\documentclass[sigconf,nonacm]{acmart}

\definecolor{linkblue}{RGB}{0,80,160}
\AtEndPreamble{\hypersetup{colorlinks=true,allcolors=linkblue}}

\usepackage{amsmath}
\DeclareMathOperator{\clamp}{clamp}
\newcommand{\dd}{\mathop{}\!\mathrm{d}}

\graphicspath{{figures/}{assets/comparison/}}
\newlength{\clipgut}
\newlength{\clipgap}
\newlength{\cliptilew}
\newcommand{\cliptile}[1]{\includegraphics[width=\cliptilew]{#1}}
\newcommand{\clipgrid}[4]{%
  \begingroup
    \setlength{\tabcolsep}{0pt}%
    \setlength{\lineskip}{0pt}%
    \setlength{\lineskiplimit}{0pt}%
    \renewcommand{\arraystretch}{0}%
    \begin{tabular}{@{}c@{\hspace{\clipgut}}c@{}}
      \cliptile{#1} & \cliptile{#2} \\[\clipgut]
      \cliptile{#3} & \cliptile{#4}
    \end{tabular}%
  \endgroup}
\newlength{\sungap}
\newlength{\sunsep}
\newlength{\suntilew}
\newlength{\fitgap}
\newlength{\fittilew}
\newlength{\cmpgap}
\newlength{\cmpsep}
\newlength{\cmptilew}
\AtBeginEnvironment{thebibliography}{\interlinepenalty=10000}

\AtBeginDocument{%
}

\begin{document}

\title[Windfoil: Closed-Form Coverage for Real-Time and Differentiable Vector Graphics]
  {\texorpdfstring{Windfoil: Closed-Form Coverage for Real-Time and\\
  Differentiable Vector Graphics}
  {Windfoil: Closed-Form Coverage for Real-Time and Differentiable Vector Graphics}}

\author{Matt DesLauriers}
\affiliation{%
  \department{Creative Computing Institute}
  \institution{University of the Arts London}
  \city{London}
  \country{United Kingdom}}
\email{m.deslauriers0620251@arts.ac.uk}

\begin{abstract}
We present Windfoil, a GPU-friendly algorithm that treats rasterisation and
differentiable vector graphics as two sides of the same problem by evaluating
the box-filtered winding number of quadratic B\'ezier contours in closed form.
We implement this in WebGPU, allowing it to run across a range of environments,
including a web browser on a consumer laptop, and apply the system to real-time
2D rendering, high-resolution rasterisation for print media, and a
differentiable renderer. We compare our renderer against Skia, a
production-grade engine, and Slug, a popular GPU rasterisation algorithm for
games and real-time applications, measuring fidelity to a reference
box-filtered coverage. Our renderer matches the reference more closely than
either, at performance comparable to Slug. We also compare our optimiser
against DiffVG and B\'ezier Splatting, where it reaches equivalent or better
reconstruction quality at a fraction of the per-step cost, scaling to tens of
thousands of shapes at interactive rates.
\end{abstract}

\begin{teaserfigure}
  \centering
  \setlength{\fitgap}{3pt}
  \setlength{\fittilew}{\dimexpr(\linewidth-4\fitgap)/5\relax}
  \begin{tabular}{@{}c@{\hspace{\fitgap}}c@{\hspace{\fitgap}}c@{\hspace{\fitgap}}c@{\hspace{\fitgap}}c@{}}
    \includegraphics[width=\fittilew]{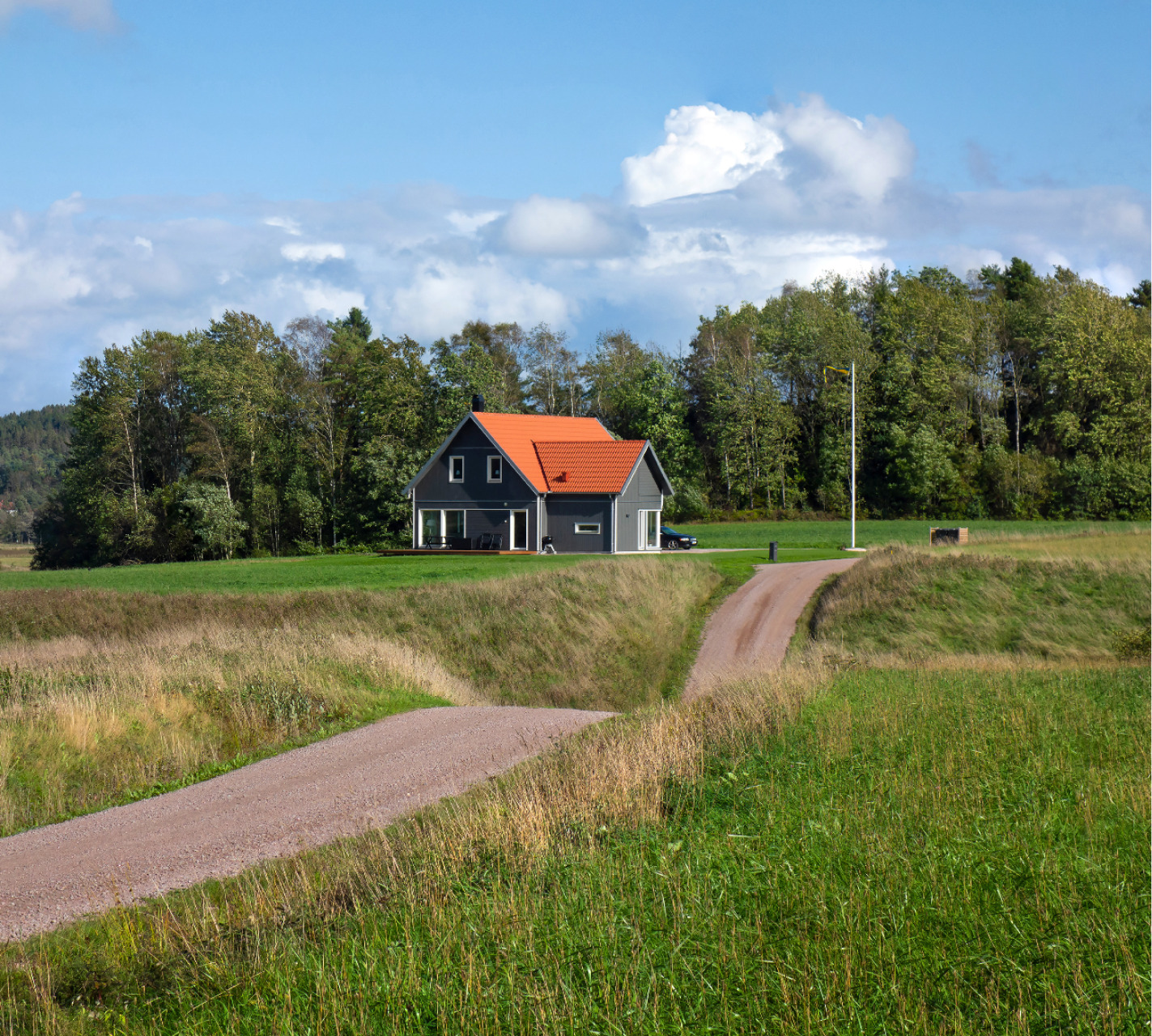} &
    \includegraphics[width=\fittilew]{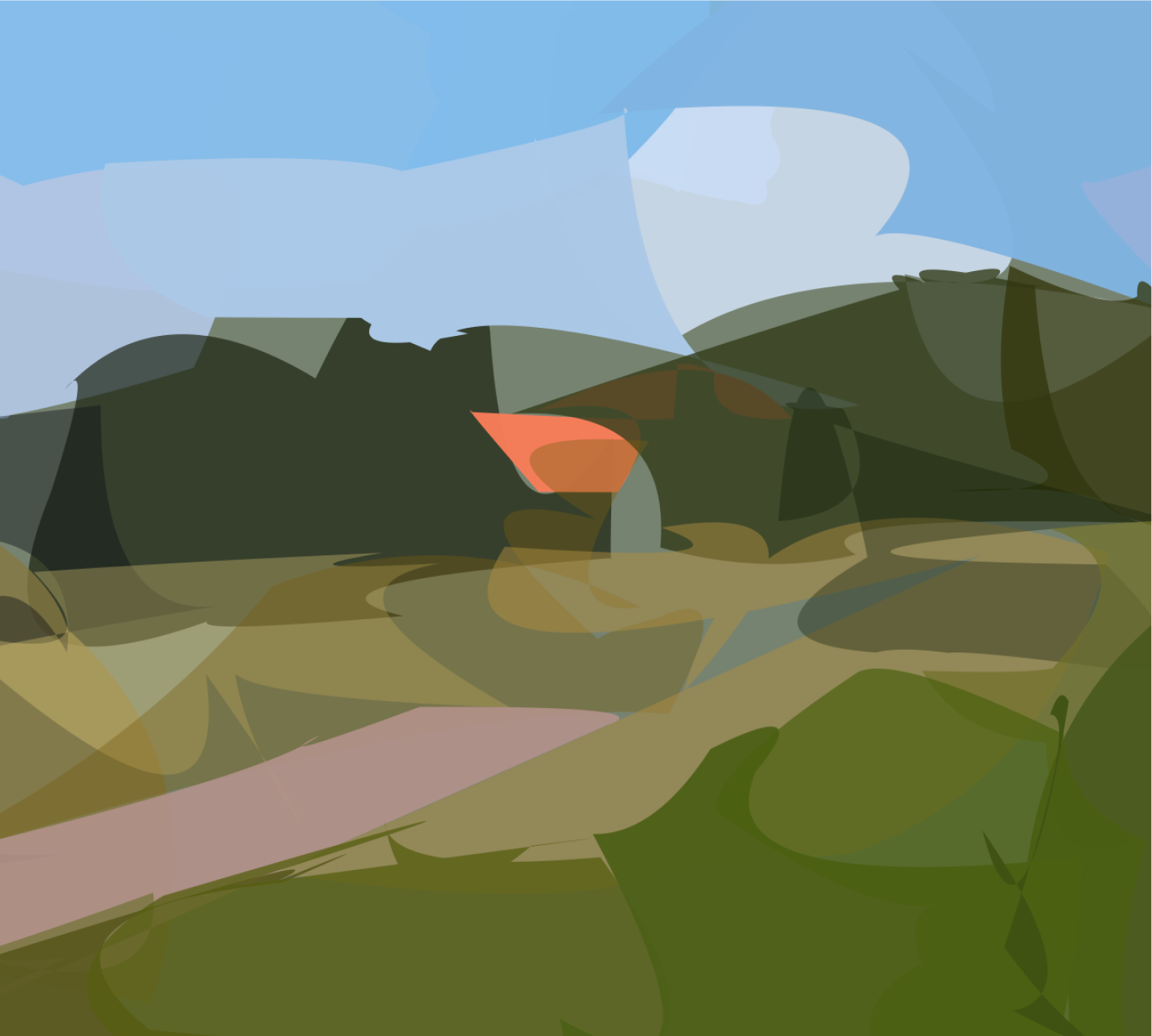}          &
    \includegraphics[width=\fittilew]{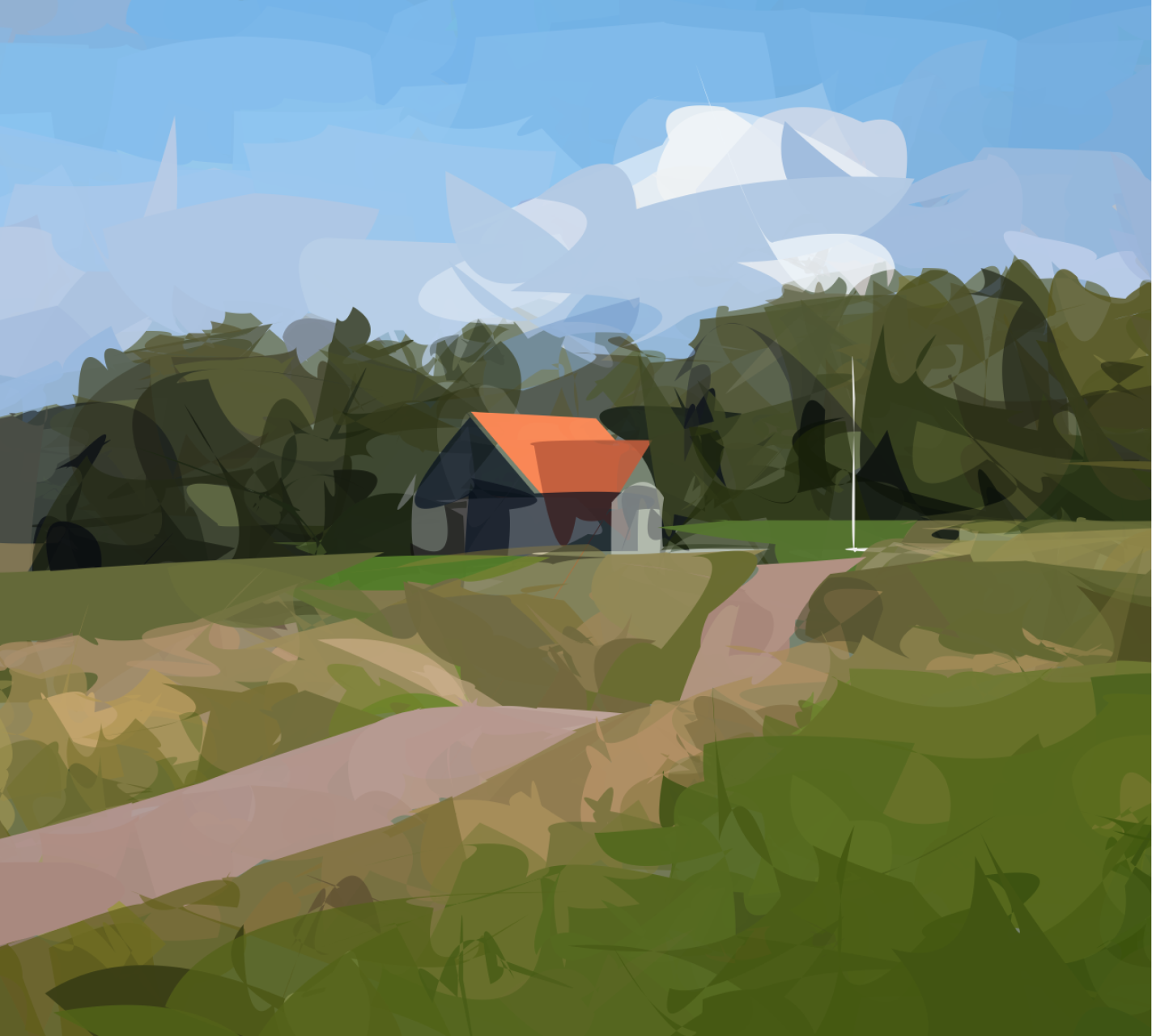}         &
    \includegraphics[width=\fittilew]{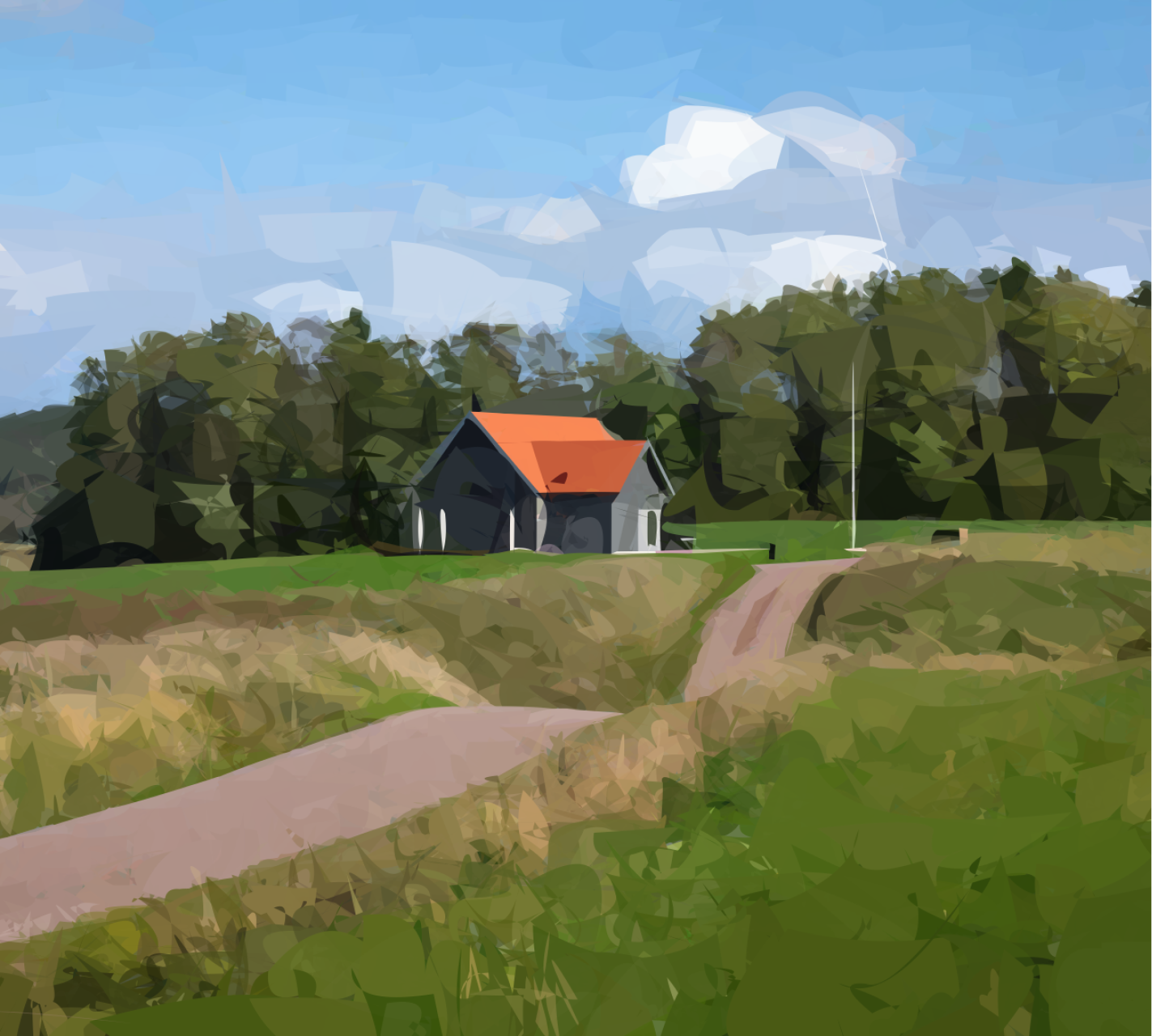}        &
    \includegraphics[width=\fittilew]{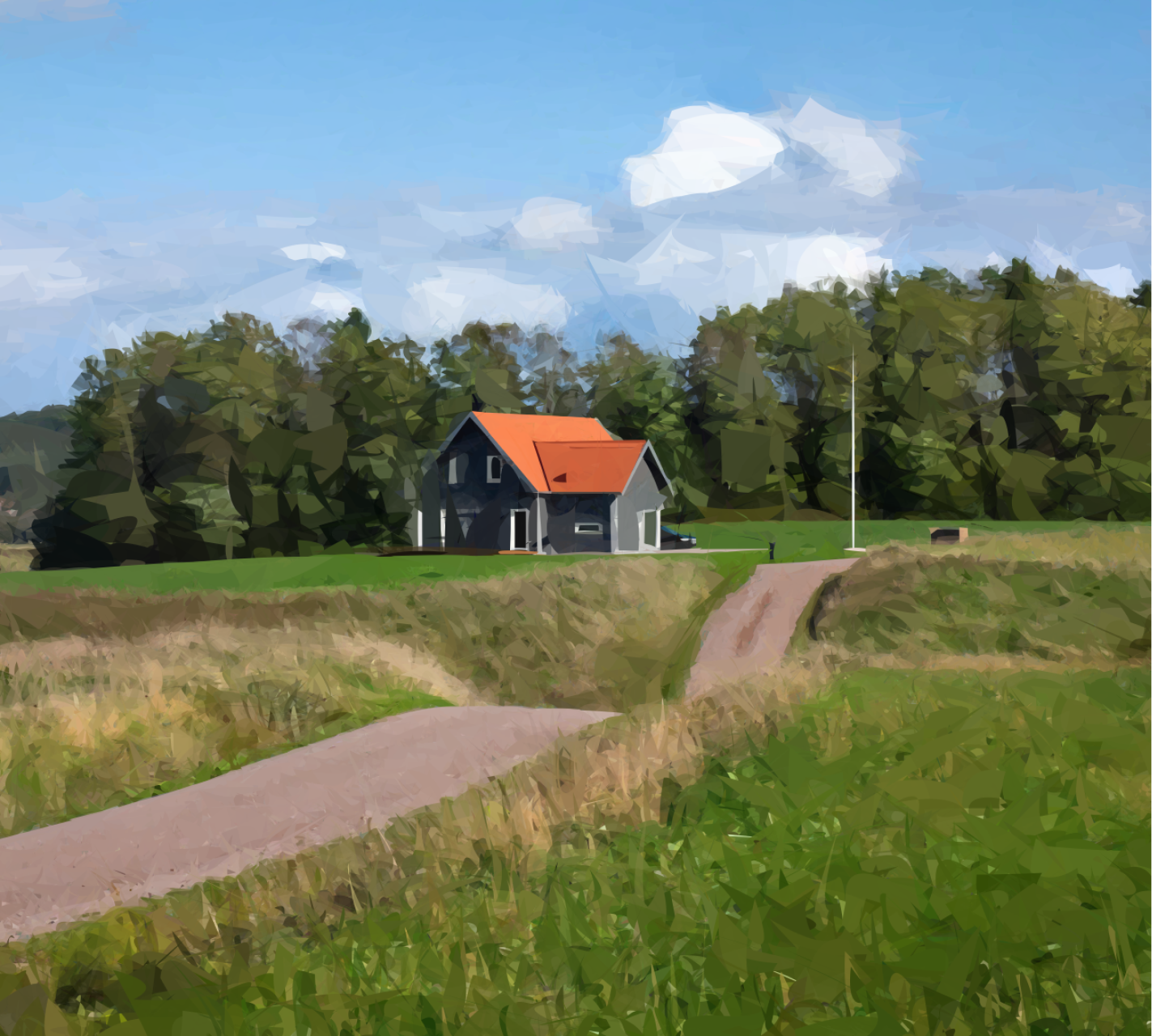}       \\[2pt]
    \footnotesize Ground Truth &
    \footnotesize $N{=}32$     &
    \footnotesize $N{=}512$    &
    \footnotesize $N{=}4{,}096$ &
    \footnotesize $N{=}50{,}000$ \\
  \end{tabular}
  \caption{Approximating a target photograph with 32 to 50,000 shapes, each a closed loop of quadratic B\'ezier curves, fitted by gradient descent on mean-squared pixel error. The renderer is differentiable and runs in WebGPU.}
  \Description{Five images in a row, the first is a photo of a house in a field, and each subsequent image shows the same scene but painted with an increasing density of abstract coloured shapes.}
  \label{fig:fitting}
\end{teaserfigure}

\maketitle

\section{Introduction}

Vector graphics has long been a central pillar of computing research. Many of
today's production graphics engines take advantage of the GPU to accelerate 2D
rendering of text, icons, emojis, shapes, and user-interface
elements~\cite{skia2026,kilgard2012path}. Acceleration has become particularly
important in interactive fields such as rendering on the web, design tools,
and games. Achieving crisp and scalable text and vector graphics at real-time rates on GPU
hardware remains challenging, particularly with regard to antialiasing. Vector shapes
are commonly rasterised with a winding rule, and applying this once per pixel produces
a binary image with visible artefacts and stepping along curved and diagonal edges. To smooth
these edges, antialiasing is used, which is often handled separately from the winding computation,
such as through multisampling, analytic coverage, or baking distance fields into a texture sampler.
Analytic antialiasing has attracted considerable interest because it can produce
legible text and smooth edges across resolutions. However, existing GPU methods often trade accuracy
or scale invariance for speed, approximating coverage or flattening curves to line segments.

Some of these problems also arise in machine learning during
\emph{differentiable rendering}, in which a renderer also provides
derivatives of the output image with respect to scene parameters, allowing these parameters to be
optimised by gradient descent. Rather than authoring a scene
directly, a differentiable renderer allows a scene's curves, colours, and
appearance to be optimised toward an objective, including raster-based losses
and losses from vision-language models. Applications include image
vectorisation and abstraction~\cite{li2020diffvg,ma2022layerwise,berio2025neural},
sketch synthesis~\cite{frans2022clipdraw,vinker2022clipasso,vinker2023clipascene,
xing2023diffsketcher}, vector
generation~\cite{reddy2021im2vec,jain2023vectorfusion,zhang2024neuralpath,
xing2024svgdreamer}, semantic typography~\cite{iluz2023wordasimage}, and
3D wire art and multi-view synthesis~\cite{qu2024wired}.

Rasterisation is not naturally differentiable: a winding test classifies a point
as inside or outside a shape, so the derivative is zero almost everywhere.
Existing differentiable renderers typically resort to Monte Carlo sampling or smooth
analytic approximations~\cite{li2020diffvg,liu2025bezier}. In practice, differentiable rendering
is used only during optimisation, and the results are exported as vector paths to be displayed
with a renderer such as Skia, which relies on a different coverage model.

Windfoil addresses these problems by computing the box-filtered winding number
analytically in closed form over a given pixel footprint, without multisampling or baked texture
storage. This evaluation can be used for real-time animation, high-resolution print export, and differentiable rendering.
Because the display and differentiable paths share the same closed-form evaluation, gradients follow the
same coverage model as the rendered image. The system runs in WebGPU within a web browser,
and can scale to more complex scenes through headless and remote rendering on the server.

The formulation presents an extensible primitive for 2D vector graphics. We
showcase it through capabilities that were previously difficult to achieve
at interactive speeds on the web, including adjustable filter kernels,
real-time shape fitting, and text-to-sketch synthesis guided by CLIP, a
vision-language model~\cite{radford2021clip}.

In summary, we contribute:

\begin{itemize}
  \item a closed-form and per-pixel solution for the box-filtered winding number of
    closed quadratic B\'ezier contours that is exact for well-formed
    shapes;
  \item a real-time WebGPU implementation built on this solution,
    demonstrated in a fragment shader for display and a compute pipeline for
    differentiable optimisation;
  \item acceleration structures tailored to each path: horizontal row binning
    for the fragment shader and tiled binning for the compute dispatch; and
  \item a renderer that matches a box-filter reference more closely than Skia
    or Slug, and an optimiser that is faster than DiffVG and B\'ezier Splatting
    at comparable or better reconstruction quality, while running across a wider
    range of consumer runtimes, including client-side in a web browser.
\end{itemize}

Code, documentation, validation, and benchmarks are available in the 
\href{https://github.com/texel-org/windfoil-paper}{reference repository} and
\href{https://github.com/texel-org/windfoil-algorithm}{companion demo}.

\section{Related Work}

\subsection{Hardware-Accelerated Vector Graphics}

A number of techniques bring hardware acceleration to text and vector
graphics~\cite{kilgard2012path}. Skia, an established production-grade 2D
rendering engine used by the Chrome browser and many other applications, offers
multiple acceleration paths for antialiased shapes, ranging from multisampling
to analytic coverage~\cite{skia2026}.

Text is a special case of vector rendering. Text glyphs often have simple,
well-formed curves and fit neatly within a rectangle, yet high-quality text
rendering poses its own challenges, including legibility across zoom levels and
support for large glyph atlases such as CJK character sets and
Unicode symbols. At fixed sizes, curves can be flattened and rasterised to a
device tolerance, but this can be costly for real-time,
scale-invariant rendering, which is important in games, design tools, maps, and
interactive media. Valve popularised baking signed
distance fields for text and decals in 3D games~\cite{green2007sdf}, and this was later
refined by Chlumsk\'y's msdfgen~\cite{chlumsky2015thesis,chlumsky2016msdfgen}. Baked
distance fields, however, can be slow to compute, memory hungry, restricted to
a fixed range of resolutions, and prone to visible artefacts on certain
curves.

Several systems instead render directly from path data on the GPU. Loop and Blinn
triangulate shapes and evaluate an implicit form of each curve per fragment~\cite{loop2005resolution}.
Nehab and Hoppe store the curves overlapping each cell of a coarse lattice and approximate coverage
by filtering the distance to the nearest curve segment~\cite{nehab2008random}. 
Dobbie's vector texture approach approximates a per-pixel area integral over a
circular footprint by averaging ray coverage at several
angles~\cite{dobbie2016vector}; a later open-source reimplementation integrates
the winding number against a one-pixel box along a horizontal and a vertical
ray~\cite{greenlightning2022gpu}. Wallace's GPU text rendering draws a triangle fan per glyph, using
additive blending in a colour buffer to accumulate winding~\cite{wallace2016text}.
Pathfinder flattens curves into line segments and
sums their signed coverage per tile on the GPU~\cite{silva2019pathfinder}. Vello's compute renderer builds
on this with a complete GPU pipeline, evaluating each segment's pixel area in closed form and
carrying winding between tiles with a prefix sum~\cite{linebender2024vello}.
Slug estimates coverage in a fragment shader from the crossings of horizontal
and vertical rays~\cite{lengyel2017slug}. It has been widely licensed in the
video game industry for rendering text and graphics, and Lengyel dedicated its
patent to the public domain in March 2026, also publishing reference shaders under an open-source licence~\cite{lengyel2026decade}.
Windfoil takes inspiration from Slug's execution model of a per-fragment
gather over bands, although with a different coverage implementation and acceleration structure.

Our approach is also closely related to prior work on exact filtering. Catmull
computed exact box-filtered coverage by clipping polygons to each pixel as early
as 1978~\cite{catmull1978hidden}. Manson and Schaefer rasterise shapes
bounded by B\'ezier curves in closed form using a boundary integral against
piecewise-polynomial filters, accumulating contributions along scanlines~\cite{manson2013analytic}.
Olson filters straight-edged polygons exactly by clipping them to each pixel and applying Green's theorem~\cite{olson2024exact}.
Ellis et al.\ decompose glyphs into trapezoids and piecewise-biquadratic
segments, evaluating trapezoid coverage exactly and approximating curved
coverage with trapezoidal quadrature within each pixel~\cite{ellis2019realtime}.

Building on these ideas, we integrate the winding number of quadratic contours in closed form, gather
contributions independently per pixel using horizontal row bands, and apply the fill rule after
integration. This organisation supports both real-time display and
differentiation. A recent, self-published preprint for text
rendering~\cite{rook2026scanline} also evaluates coverage independently per
pixel, clipping quadratic curves to the pixel footprint but approximating the
clipped area with rectangles and trapezoids rather than integrating it exactly.

\subsection{Differentiable Vector Graphics}

DiffVG introduced much of the current discourse on differentiable vector
graphics in 2020 and offers two render modes: one based on Monte Carlo sampling
and another using an analytic approximation~\cite{li2020diffvg}. B\'ezier
Splatting~\cite{liu2025bezier} renders curves as many Gaussians using an approach
similar to Gaussian splatting, enabling a significant performance boost over DiffVG for certain tasks.

A substantial body of applications builds on these renderers, including work
broadly described as text-to-sketch or text-to-vector synthesis: CLIPDraw,
CLIPasso, CLIPascene, DiffSketcher, and
SVGDreamer~\cite{frans2022clipdraw,vinker2022clipasso,vinker2023clipascene,
xing2023diffsketcher,xing2024svgdreamer}. Each optimises through the underlying differentiable vector
renderer, so improvements in performance and reconstruction quality can directly benefit these systems.

\section{Method}

We represent 2D shapes as one or more closed, directed contours of quadratic
B\'ezier curves. A point's winding number is the signed count of boundary
crossings along a horizontal ray cast to the right. A nonzero or even-odd fill
rule uses this count to determine whether the point is inside the shape.

To produce antialiased pixels, we average the winding number over a rectangular
footprint and then apply the fill rule. We compute this average from the
boundary curves, evaluating each pixel independently.

\subsection{Box-Filtered Winding}

Let \(p=(p_x,p_y)\) be a pixel's centre in curve coordinates. We average
the winding number over a rectangular footprint around \(p\),
\begin{equation}
  B(p,s)=[p_x-h_x,\,p_x+h_x]\times[p_y-h_y,\,p_y+h_y],
  \label{eq:footprint}
\end{equation}
where \(s=(s_x,s_y)\) is the filter size in curve coordinates, normally
the size of one pixel, and \(h_x=s_x/2\), \(h_y=s_y/2\) are its
half-width and half-height. Increasing \(s\) averages over a larger area
and gives softer edges. We use an axis-aligned box, which is exact during translation
and scaling, but an approximation under rotation, shear, or perspective transform.

Let \(w(x,y)\) be the winding number at a point \((x,y)\). The box-filtered winding is
the winding integral over the footprint divided by the footprint area:
\begin{equation}
  F(p,s)
  =
  \frac{1}{s_xs_y}
  \iint_{B(p,s)} w(x,y)\,\dd A.
  \label{eq:filtered-winding}
\end{equation}
A fill rule then maps \(F\) into a coverage value in \([0,1]\).
For the nonzero rule,
\begin{equation}
  C_{\mathrm{nz}}(F)=\min\!\left(\lvert F\rvert,\,1\right),
  \label{eq:nonzero-coverage}
\end{equation}
and for the even-odd rule,
\begin{equation}
  C_{\mathrm{eo}}(F)=\min\!\left(r,\,2-r\right),
  \qquad
  r=\lvert F\rvert \bmod 2,
  \label{eq:evenodd-coverage}
\end{equation}
which gives the distance from \(F\) to the nearest even integer.
Both rules are independent of the winding sign convention.

\begin{figure*}[t]
  \centering
  \setlength{\cmpgap}{1.5pt}
  \setlength{\cmpsep}{6pt}
  \setlength{\cmptilew}{\dimexpr(\linewidth-5\cmpgap-\cmpsep)/7\relax}
  \setlength{\lineskip}{0pt}
  \begin{tabular}{@{}c@{\hspace{\cmpgap}}c@{\hspace{\cmpgap}}c@{\hspace{\cmpsep}}c@{\hspace{\cmpgap}}c@{\hspace{\cmpgap}}c@{\hspace{\cmpgap}}c@{}}
    \includegraphics[width=\cmptilew]{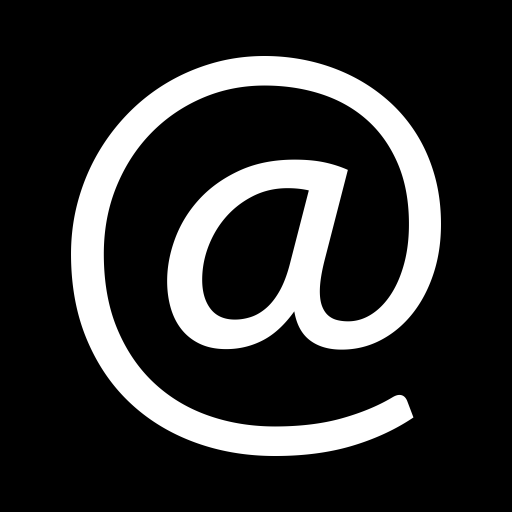}        &
    \includegraphics[width=\cmptilew]{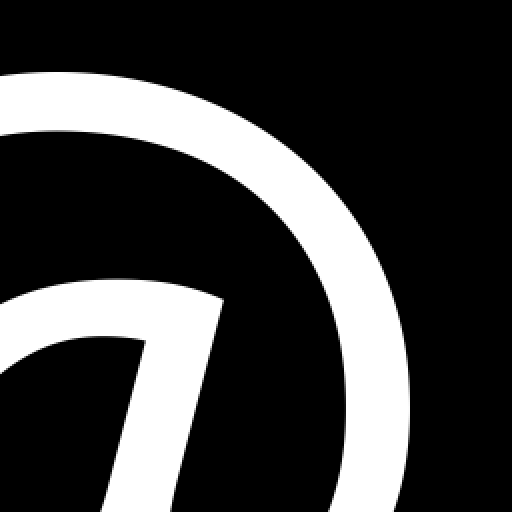}   &
    \includegraphics[width=\cmptilew]{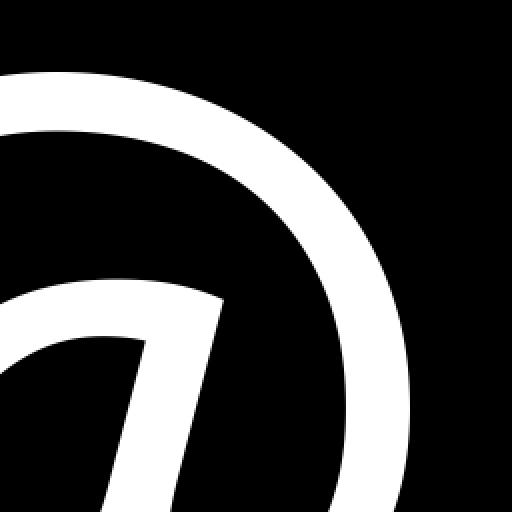}      &
    \includegraphics[width=\cmptilew]{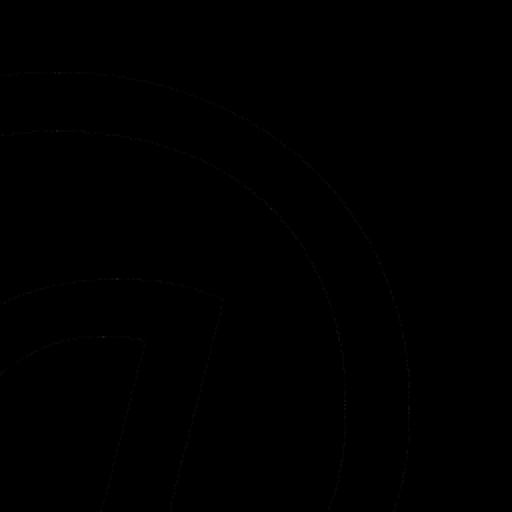}      &
    \includegraphics[width=\cmptilew]{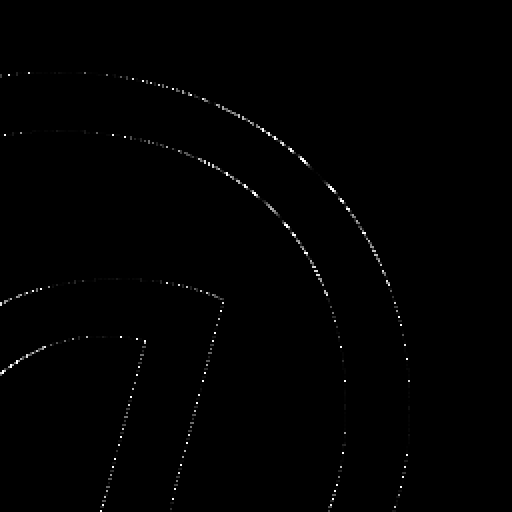}    &
    \includegraphics[width=\cmptilew]{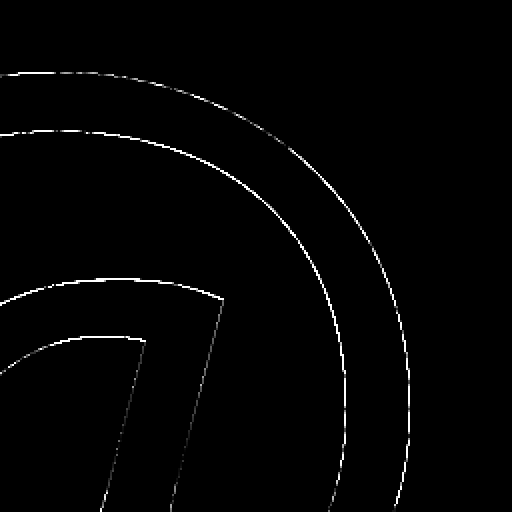}  &
    \includegraphics[width=\cmptilew]{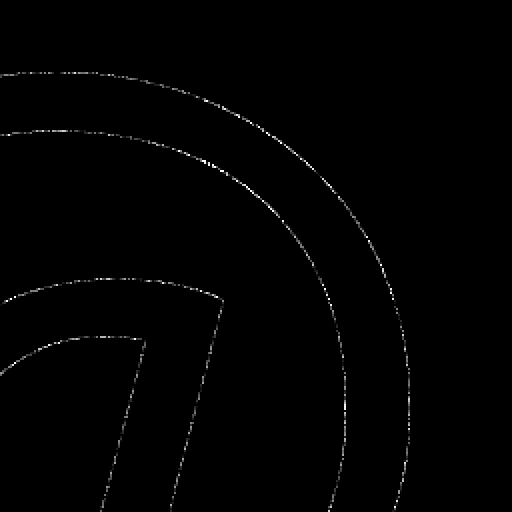}   \\[\cmpgap]
    \includegraphics[width=\cmptilew]{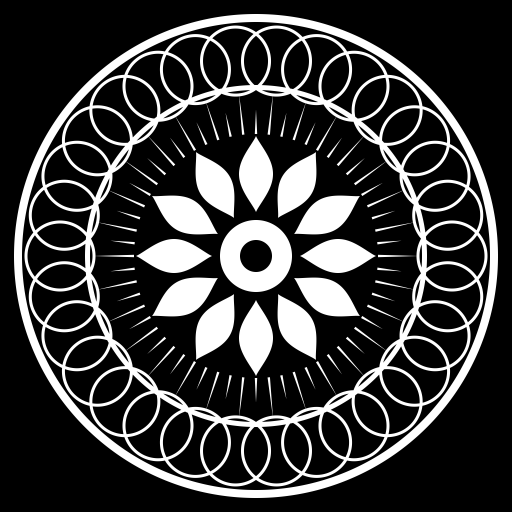}       &
    \includegraphics[width=\cmptilew]{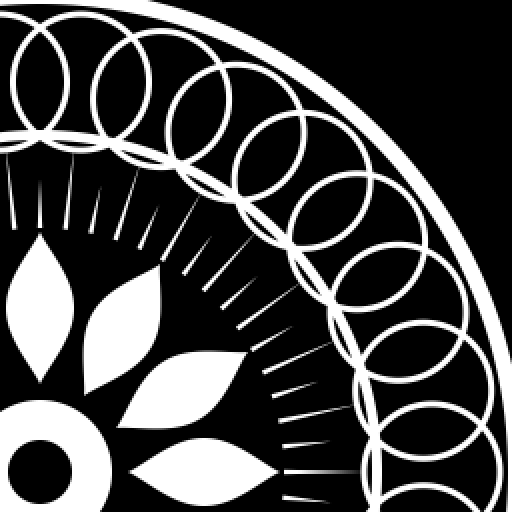}  &
    \includegraphics[width=\cmptilew]{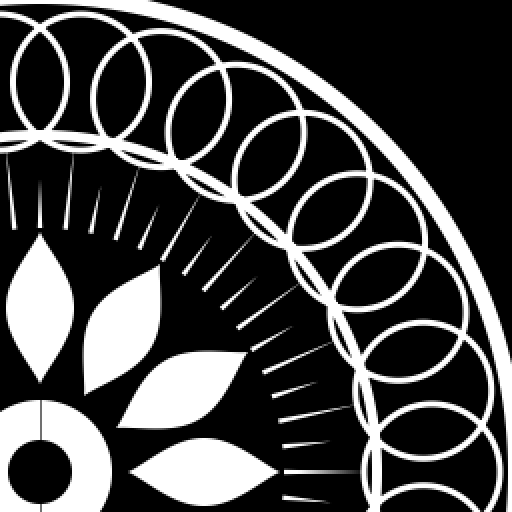}     &
    \includegraphics[width=\cmptilew]{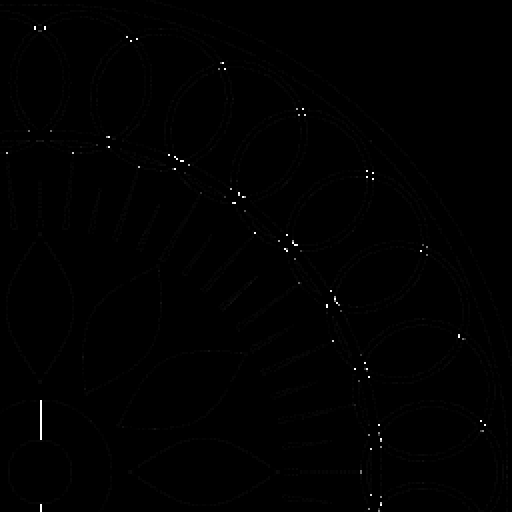}     &
    \includegraphics[width=\cmptilew]{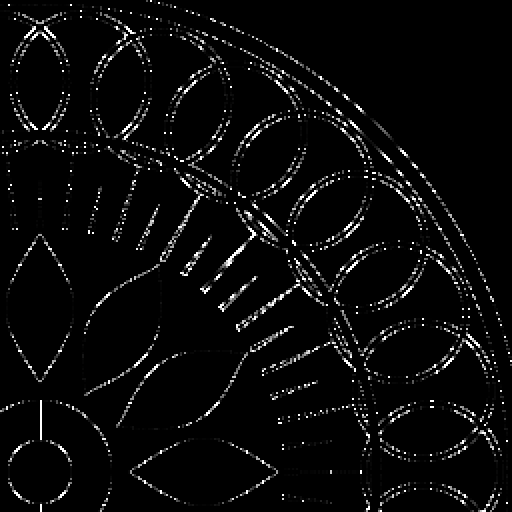}   &
    \includegraphics[width=\cmptilew]{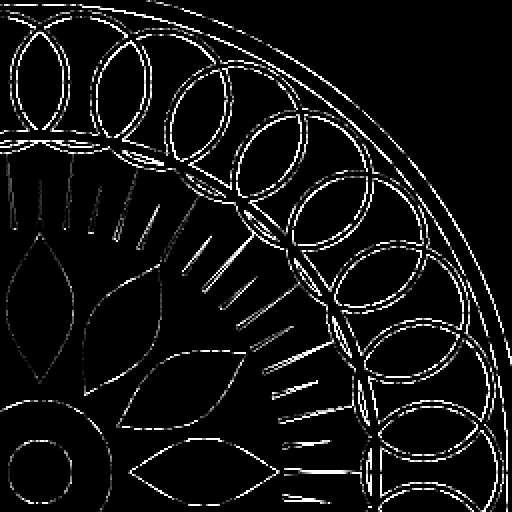} &
    \includegraphics[width=\cmptilew]{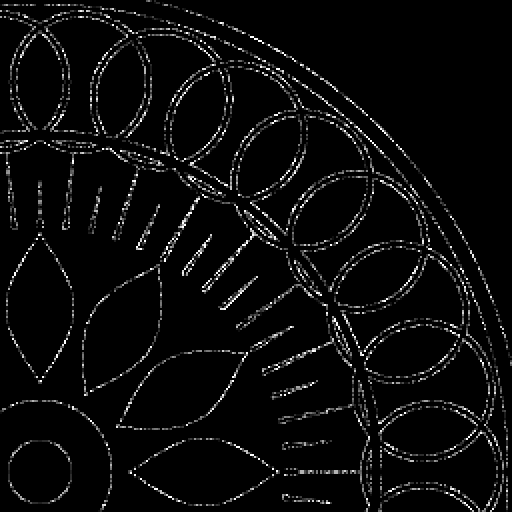}  \\[2.5pt]
    \scriptsize (a) Ground Truth  &
    \scriptsize (b) Ground Truth $2\times$ &
    \scriptsize (c) Windfoil $2\times$ &
    \scriptsize (d) $|\Delta|$ Windfoil &
    \scriptsize (e) $|\Delta|$ Slug     &
    \scriptsize (f) $|\Delta|$ Skia     &
    \scriptsize (g) $|\Delta|$ Figma    \\
  \end{tabular}
  \caption{Coverage error against the box filter. Two scenes rendered at $512\times512$ pixels: 
  the `@' glyph from the Lato font, and a synthetic SVG rosette designed to highlight winding failures.
    (a)~The box-filtered Ground Truth reference is estimated by point-sampling a $64\times64$ grid per pixel with
    winding ray-cast from the raw curves. (b)~A nearest-neighbour scaled crop of the reference image.
    (c)~Rendered with Windfoil, then also scaled and cropped. (d)--(g)~Per-pixel
    $|\Delta|$ from the reference crop as rendered by: Windfoil, Slug, Skia and Figma (v126.7.10), respectively. 
    The error is amplified $\times15$ and clipped, so white pixels show differences.
    The (d) Windfoil column is mostly black, as the majority of rendered edges agree with the box filter. The rosette
    graphic exposes areas where averaging the winding before applying the fill rule leads to failures.
    Most notably, a bright vertical seam from winding cancellation at a self-crossing contour renders incorrectly as a hole
    with Windfoil and Slug, while Skia and Figma rasterise this edge correctly.}
  \Description{A grid of two rows and seven columns. The top row is the glyph at-sign,
    the bottom a rosette graphic. The first column is the ground-truth
    reference, the next two are magnified crops of the reference and of
    Windfoil, and the last four are amplified error maps for Windfoil, Slug,
    Skia and Figma. Windfoil has smaller errors along ordinary edges, but
    both Windfoil and Slug show a bright seam at the rosette's self-crossing.}
  \label{fig:coverage}
\end{figure*}

To evaluate curve contributions independently, we apply the fill rule after
averaging the winding, whereas exact box-filtered coverage applies it at each point. 
Both agree exactly wherever the winding varies by at most one across the
footprint, such as along an isolated edge, but can differ at self-crossings and
overlapping contours, where it varies by more than one (Figure~\ref{fig:coverage}). Errors are confined to 
the pixels whose footprints intersect these regions, a limitation also shared by
other GPU renderers such as Slug and Vello.

\subsection{Boundary Integral}

By Green's theorem, the integral over the footprint in Equation~\eqref{eq:filtered-winding} can
be rewritten as a boundary integral, summed along each boundary curve. We centre the
footprint at the origin, so that it is the box \(B=[-h_x,h_x]\times[-h_y,h_y]\).

Consider a horizontal strip of the footprint at height \(y\), and a curve
crossing it at horizontal position \(x\). The width of the strip between
the footprint's left edge and the curve is
\begin{equation}
  G(x)=\clamp(x+h_x,\,0,\,s_x).
  \label{eq:clipped-width}
\end{equation}
Integrating \(G(x)\,\dd y\) along every contour within the footprint's vertical
range yields the winding integral, with rising crossings adding their clipped width
and falling crossings subtracting it.

During CPU preprocessing, we subdivide each quadratic curve at its
\(x\)- and \(y\)-extrema into at most three pieces that are monotone in
both coordinates. Per pixel on the GPU, we clip each piece to the
footprint's vertical range and find where it crosses the footprint's left
and right edges by solving a quadratic for each. As the pieces are monotone,
they cross each edge at most once, leaving up to three parts. A part left of the
footprint contributes nothing, a part to the right contributes the full width \(s_x\) times
the height it spans, and the part inside contributes a polynomial that we integrate in closed form.

\subsection{Rasterisation and Differentiation}

We sum the boundary contributions to obtain the winding integral, then divide by the footprint
area to obtain the box-filtered winding \(F\). The selected fill rule maps \(F\) to a coverage
value, which we multiply by the shape's opacity. We then combine shapes in painter's order using
standard source-over compositing.

With positive filter sizes, coverage changes continuously as the curves move, and our evaluation
is piecewise differentiable. We differentiate to obtain gradients with respect to control points,
colours, and opacities. Where derivatives are undefined or numerically unstable, we use fixed
derivative conventions and numerical safeguards.

\section{GPU Acceleration}

Windfoil targets JavaScript runtimes and WebGPU, so the same code runs
client-side in a web browser and server-side through Node.js (Dawn) or Deno
(built-in wgpu or Dawn). We implement two backends, each accelerated
for a different use case. In the display backend, curve preprocessing is performed once per
unique shape. In the differentiable backend, the preprocessing must be 
re-evaluated in each optimisation step as the geometry changes. In both cases,
curve preprocessing occurs on the CPU, and this is an aspect of the renderer that
could eventually be improved and implemented in a GPU compute pipeline.

\subsection{Rasterisation for Display}

For real-time display, we draw shapes as instanced quads with standard painter-order compositing
and alpha blending. Each quad is padded in the vertex shader to provide a sufficiently wide
boundary for antialiased coverage. Curve data is uploaded once per unique shape and shared across
instances, so thousands of repeated shapes such as text characters can
reference the same curve data. Each fragment reads only the curves belonging to
its own shape.

To avoid looping over every curve in the shape, the monotone pieces are filed
into horizontal row bands spanning the shape's vertical extent, and a fragment
reads only the bands touched by its footprint's vertical slab. Because the integral
sweeps horizontally, one row-band structure holds everything a fragment needs for coverage,
unlike Slug's dual-ray model, which must store both horizontal and vertical bands.
For each monotone piece, the shader only evaluates bands overlapping the current pixel footprint.

During pre-processing, bands with more than four pieces are sorted from right to left by the largest
\(x\) coordinate of their control points. In the shader, traversal can stop once all remaining
pieces lie to the left of the footprint. Shorter bands are left unsorted and traversed fully.
Perfectly horizontal pieces have \(y'(t)=0\) and therefore contribute zero to the winding integral;
the display backend can skip these.

The real-time rendering path also includes a minification guard. As shapes are
minified, the pixel footprint encompasses more curves and row binning becomes
underutilised. Beyond a threshold, roughly when a shape's bounding box falls below
four pixels in width and height, the shader uses a precomputed banded ink profile that
approximates its coverage.

\subsection{Differentiable Coverage}

As the algorithm is analytic and closed form, we can also extend this approach to differentiable rendering.
Our first implementation was a JAX port~\cite{jax2018}, demonstrating that automatic differentiation can produce
gradients with respect to control points, colour, opacity, and filter width.
For differentiable rendering, each curve produces up to three candidate slots
for monotone pieces so that the evaluation has a uniform upper bound;
zero-length intervals and collapsed pieces contribute nothing and may be
omitted by GPU packing.

\subsection{Accelerating Differentiation}

Our JAX implementation serves as a reference for correctness rather than as an efficient rendering path,
and we use it primarily to validate our WebGPU-accelerated implementation.

Curves are preprocessed on the CPU and passed to WebGPU compute pipelines for differentiable rendering and evaluation.
Unlike the display path, compute shaders cannot use instanced quads to restrict fragment invocations to each shape's bounds.
Instead, we tile the scene and construct lists of potentially contributing shapes for each tile, in draw order.
The forward pass walks this list for each pixel, evaluates the box-filtered winding, and composites the
shapes in painter's order.

For image fitting, we compare the rendered \(H\times W\) image \(I\) with a
target \(I^\star\) on the GPU using the mean-squared error over pixels \(i\)
and colour channels \(c\):
\begin{equation}
  \begin{aligned}
    \mathcal{L}(I,I^\star)
      &=\frac{1}{3HW}\sum_{i,c}
        \left(I_{i,c}-I^\star_{i,c}\right)^2,\\
    \frac{\partial\mathcal{L}}{\partial I_{i,c}}
      &=\frac{2}{3HW}\left(I_{i,c}-I^\star_{i,c}\right).
  \end{aligned}
  \label{eq:mse-loss}
\end{equation}
The image-space gradient is passed as the cotangent to the backward pass. We
can also support non-WebGPU differentiable objectives that provide an
image-space gradient, such as a Python server computing CLIP loss between the
rendered image and a text prompt.

The backward pass is an explicit analytic vector-Jacobian product
(VJP) on the GPU. It walks the same tile lists, recomputing the forward quantities,
and propagates the image gradient back through compositing, opacity, the fill rule, and
the winding integral. Gradients are summed over pixels, then mapped from the monotone pieces
back to the original quadratic control points.

The renderer returns gradients with respect to decoded curves, colours, and
opacities. These are not always the parameters being optimised directly; a model may
instead parameterise a shape by a centre and radius, a transformation, or
another representation. A final model-specific pullback maps the
renderer gradients onto these underlying parameters, before an Adam optimiser
step~\cite{kingma2015adam}.

\section{Results}

\subsection{Rasterisation}

\begin{figure*}[t]
  \centering
  \setlength{\sungap}{0pt}
  \setlength{\sunsep}{0pt}
  \setlength{\suntilew}{\dimexpr(\linewidth-4\sungap-\sunsep)/6\relax}
  \begin{tabular}{@{}c@{\hspace{\sungap}}c@{\hspace{\sungap}}c@{\hspace{\sunsep}}c@{\hspace{\sungap}}c@{\hspace{\sungap}}c@{}}
    \includegraphics[width=\suntilew]{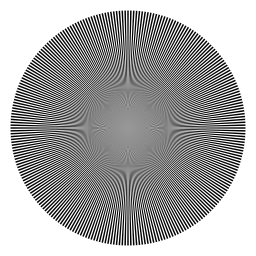}    &
    \includegraphics[width=\suntilew]{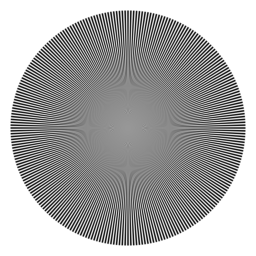}   &
    \includegraphics[width=\suntilew]{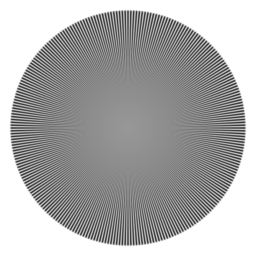}  &
    \includegraphics[width=\suntilew]{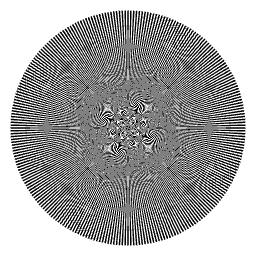} &
    \includegraphics[width=\suntilew]{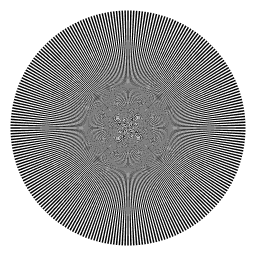}  &
    \includegraphics[width=\suntilew]{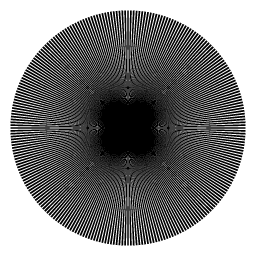}     \\[2pt]
    \footnotesize\shortstack{Windfoil\\(box)}             &
    \footnotesize\shortstack{Windfoil\\(tent)}            &
    \footnotesize\shortstack{Windfoil\\(cubic B-spline)}  &
    \footnotesize Skia &
    \footnotesize Figma         &
    \footnotesize Photoshop     \\
  \end{tabular}
  \caption{A sunburst scene of 256 thin strokes, rendered with different filter kernels and compared to existing production renderers.}
  \Description{Six renderings of the same radial sunburst of thin black rays on
    white. The first three, rendered by Windfoil with box, tent, and cubic
    B-spline filters, resolve toward a smooth uniform grey at the centre. The
    last three, from Skia, Figma, and Photoshop, show strong moir\'e
    interference and a progressively darker core.}
  \label{fig:kernels}
\end{figure*}

\paragraph{Validation.}
We validate correctness against a supersampled box-filter reference and against
two widely used rendering systems: Skia, a production-grade rasterisation
engine, and Slug, a real-time shader-based algorithm designed for text
rendering (Figure~\ref{fig:coverage}).
Our reference is computed by box-downsampling a binary rasterisation of the
original curves at \(24\times\) resolution. For a controlled comparison, we
use our own WebGPU implementation of Slug based on its reference
code~\cite{lengyel2017slug,lengyel2026slugcode}. Skia is accessed through
\texttt{@napi-rs/canvas}, a Node.js binding for Skia with an HTML5 Canvas2D compatible API.
The minification guard is disabled for this validation, so these measurements use coverage
computed from the box-filtered winding.

Across our test corpus, Windfoil's mean absolute coverage error is
\(1.2\times10^{-4}\), which is in the same range as the sampling error of the 
supersampled reference itself. By comparison, Slug and Skia in this evaluation have mean
errors of \(1.1\times10^{-3}\) and \(1.6\times10^{-3}\), respectively, about \(9\times\) and \(13\times\) larger.

Our rasterisation can disagree with the box-filtered fill at self-crossings,
overlapping contours, or multiple even-odd transitions within a footprint.
This limitation is confined to pixels whose footprint spans the conflicting winding regions.
In many cases it can be reduced with additional per-fragment samples or by
preprocessing the input contours. Skia resolves the self-crossing seam in
Figure~\ref{fig:coverage}, while Slug reproduces it, suggesting that
the trade-off may be acceptable for some real-time applications.

\paragraph{Performance.}
We benchmark per-frame time against Slug over a range of zoom levels on three scenes:
a dense grid of real text; a complex SVG (the Ghostscript tiger:
304 overlapping shapes, approximately \(22{,}000\) quadratics, and
painter-order overdraw); and a synthetic stress shape (a self-intersecting
starburst of approximately 240 quadratics whose curves pile into every pixel).

Overall performance is comparable, with our system outperforming Slug at both extremes
of zoom. Under magnification, we are \(1.2\text{--}2.6\times\)
faster, as the pixel footprint collapses to a single band and our single band axis
reads approximately half as many curves as Slug's dual ray approach. Under
minification, once a shape's ink box falls below approximately four device
pixels, our shader switches to a precomputed banded ink profile and renders
these cases \(5\text{--}6\times\) faster. Through the middle of the zoom
range, Slug leads by \(1.2\text{--}2.8\times\), depending on the scene and zoom
level. On an Apple M2 MacBook Air, both methods remain within a 60\,fps frame
budget at every zoom level for common workloads such as thousands of on-screen
text or icon glyphs.

\begin{table*}[t]
  \caption{Image fitting on an NVIDIA RTX 2000 Ada. Final PSNR is scored from
    saved PNGs, with DiffVG's final render at \(8\times8\) samples per pixel.
    $N$ is the initial shape count; $\Delta$ is Windfoil's PSNR minus the
    baseline's. Optimise time excludes
    setup, warm-up, final rendering, scoring, and output. \emph{Parity speed-up}
    divides the baseline's optimise time by Windfoil's time to first reach
    that baseline's final scored PSNR. Dashes mark unavailable or unreported
    comparisons. Bold marks the best value
    at the displayed precision, without implying statistical significance.}
  \label{tab:fitting}
  \footnotesize
  \setlength{\tabcolsep}{3.5pt}
  \begin{tabular}{@{}llrl rrr rr rrr rr@{}}
    \toprule
    & & & & \multicolumn{3}{c}{final PSNR (dB)}
      & \multicolumn{2}{c}{$\Delta$ PSNR (dB)}
      & \multicolumn{3}{c}{optimise time (s)}
      & \multicolumn{2}{c@{}}{parity speed-up} \\
    \cmidrule(lr){5-7}\cmidrule(lr){8-9}\cmidrule(lr){10-12}\cmidrule(l){13-14}
    Benchmark & Opt.\ size & $N$ & Budget & Windfoil & DiffVG & B\'ezier
      & DiffVG & B\'ezier & Windfoil & DiffVG & B\'ezier
      & DiffVG & B\'ezier \\
    \midrule
    \multicolumn{14}{@{}l}{\emph{Equal step budget}} \\
    Kodak (24) & $768\times512$ & 512 & 800 steps & \textbf{26.60} & 26.41 & 22.80 & $+0.19$ & $+3.80$ & \textbf{2.57} & 269 & 23.0 & $243\times^{\dagger}$ & $196\times$ \\
    \emph{F\"arlev} & $512\times288$ & 256 & 800 steps & \textbf{21.24} & \textbf{21.24} & 19.80 & $0.00$ & $+1.44$ & \textbf{1.22} & 138 & 17.2 & --- & $224\times$ \\
    \emph{F\"arlev} & $512\times288$ & 512 & 800 steps & \textbf{21.83} & 21.79 & 20.73 & $+0.04$ & $+1.10$ & \textbf{1.85} & 220 & 22.3 & $182\times$ & $177\times$ \\
    \emph{F\"arlev} & $512\times288$ & 4096 & 800 steps & \textbf{26.77} & 21.31 & 22.61 & $+5.46$ & $+4.16$ & \textbf{10.1} & 1658 & 132 & $4901\times$ & $214\times$ \\
    \emph{F\"arlev} & $4096\times2304$ & 256 & 800 steps & \textbf{20.93} & 20.92 & 19.55 & $+0.01$ & $+1.38$ & \textbf{17.2} & 1510 & 51.3 & $121\times$ & $151\times$ \\
    \addlinespace
    \multicolumn{14}{@{}l}{\emph{Equal optimisation-time budget}} \\
    \emph{F\"arlev} & $4096\times2304$ & 256 & 60 s$^{\ddagger}$ & \textbf{20.96} & 19.96 & 19.77 & $+1.00$ & $+1.20$ & 60.0 & 61.5 & 60.1 & --- & --- \\
    \addlinespace
    \multicolumn{14}{@{}l}{\emph{B\'ezier Splatting's published 10{,}000-step schedule}} \\
    \emph{F\"arlev} & $512\times288$ & 512 & 10k steps & \textbf{21.93} & --- & 21.40 & --- & $+0.53$ & \textbf{25.0} & --- & 295 & --- & $947\times$ \\
    \emph{F\"arlev} & $512\times288$ & 4096 & 10k steps & \textbf{27.88} & --- & 23.02 & --- & $+4.86$ & \textbf{149} & --- & 1683 & --- & $2145\times$ \\
    \addlinespace
    \multicolumn{14}{@{}l}{\emph{Large $N$ (Windfoil only)}} \\
    \emph{F\"arlev} & $4096\times2304$ & 10000 & 800 steps & 23.23 & --- & --- & --- & --- & 65.6 & --- & --- & --- & --- \\
    \emph{F\"arlev} & $4096\times2304$ & 50000 & 800 steps & 25.46 & --- & --- & --- & --- & 176 & --- & --- & --- & --- \\
    \bottomrule
  \end{tabular}
  \smallskip
  \parbox{\textwidth}{\scriptsize
    Kodak PSNR and times are means over all 24 images; parity ratios are medians.
    $^{\dagger}$\,DiffVG parity is reached on 20 of 24 images; B\'ezier parity on all 24.
    $^{\ddagger}$\,Time is measured from the first optimisation step, finishing
    the step in progress at the deadline: 2,421 steps for Windfoil, 34 for
    DiffVG, and 893 for B\'ezier. Parity is reported for fixed-step budgets only.}
\end{table*}

\subsection{Differentiation}
\label{sec:results-differentiation}

We evaluate image fitting against DiffVG~\cite{li2020diffvg} and B\'ezier
Splatting~\cite{liu2025bezier}, optimising closed shapes toward a target image
using the mean-squared pixel error in Equation~\eqref{eq:mse-loss}.

Windfoil and DiffVG share the same initial geometry, colours, and opacities,
with eight quadratic segments per shape. B\'ezier Splatting uses its native
initialisation and curve representation. Windfoil uses a one-pixel box filter
and independently optimised RGB and alpha throughout. We score each final
render with a shared evaluation of
\(\mathrm{PSNR}=-10\log_{10}(\mathrm{MSE})\).
DiffVG optimises at \(2\times2\) samples per pixel and renders its scored
image at \(8\times8\).

Benchmarks are tested on an NVIDIA RTX 2000 Ada Generation GPU (16\,GB).
Windfoil runs in Node.js with Dawn using Vulkan; both baselines use PyTorch
and CUDA. Optimisation times in Table~\ref{tab:fitting} exclude setup,
warm-up, final rendering, scoring, and file output. Targets are the 24-image
Kodak dataset~\cite{franzen1999kodak} and \emph{A dip in the road in
F\"arlev}~\cite{carter2018farlev}, referred to as \emph{F\"arlev}.

\begin{figure}[t]
  \centering
  \includegraphics[width=\linewidth]{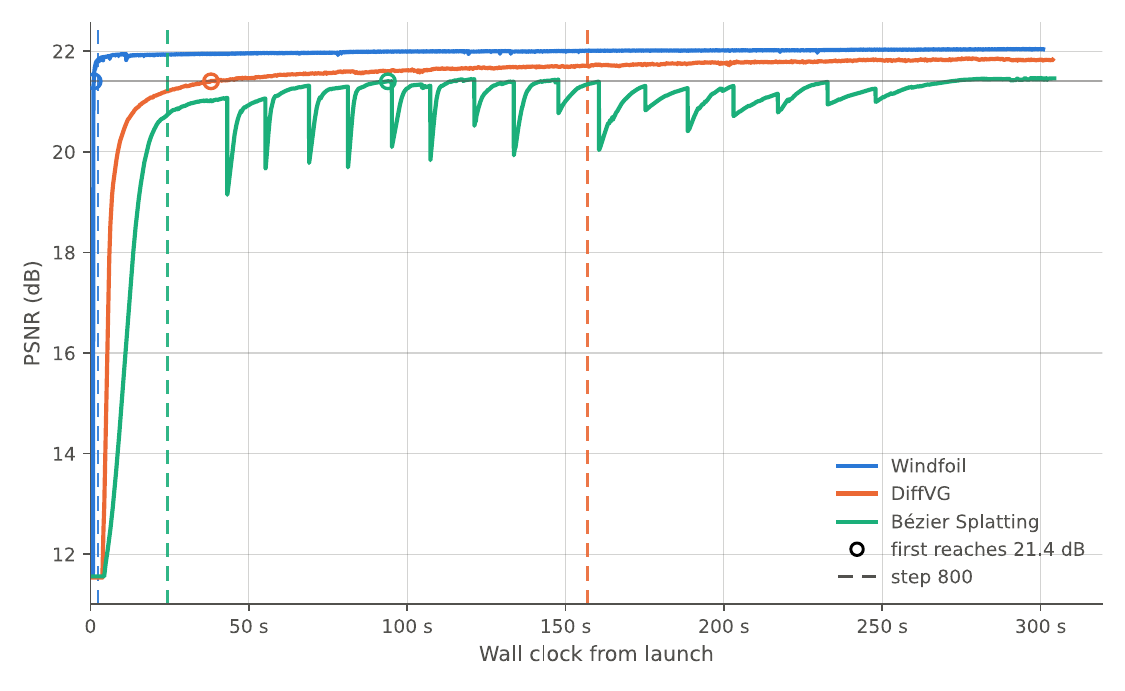}
  \caption{Convergence on \emph{F\"arlev} (\(512\times288\), \(N=512\))
    with a 300\,s optimisation budget per engine on an NVIDIA RTX 2000 Ada.
    Curves show PSNR from each engine's training loss. Time is measured from
    process launch and includes start-up (0.8\,s for Windfoil, 3.6\,s for
    DiffVG, and 4.3\,s for B\'ezier Splatting). Circles mark the first crossing of 21.4\,dB:
    1.0\,s for Windfoil, 38\,s for DiffVG, and 94\,s for B\'ezier Splatting.
    B\'ezier's repeated drops and recoveries reflect pruning and densification.
    Dashed lines mark the 800th step. Windfoil completes 115k steps and ends
    at 22.0\,dB; DiffVG completes 1.5k at 21.8\,dB, and B\'ezier Splatting
    10.8k at 21.5\,dB.}
  \Description{A line plot of PSNR in decibels against wall-clock time on a
    linear axis from zero to 300 seconds. Windfoil rises almost vertically to
    about 22\,dB within the first few seconds and stays flat. DiffVG climbs over
    the first 38 seconds to about 21.4\,dB and creeps up to 21.8\,dB. B\'ezier
    Splatting climbs more slowly and shows a repeating sawtooth of drops and
    recoveries before settling near 21.5\,dB. Circles and dashed vertical lines
    mark the first crossing of 21.4\,dB and the 800th step for each engine.}
  \label{fig:convergence}
\end{figure}

Windfoil reaches comparable or better reconstruction quality at substantially
lower optimisation cost (Table~\ref{tab:fitting} and
Figure~\ref{fig:convergence}). On Kodak at native resolution, after 800 steps with
512 shapes, mean PSNR is 26.60\,dB for Windfoil, 26.41\,dB for DiffVG, and 22.80\,dB for
B\'ezier Splatting. Mean step times are 3.2\,ms, 336\,ms, and 29\,ms,
respectively. Under this configuration, Windfoil is able to run at real-time rates, including
client-side in a web browser.

We report \emph{parity speed-up}: the baseline's total optimisation time
divided by the time Windfoil takes to reach the baseline's final scored PSNR.
On Kodak, Windfoil reaches or exceeds DiffVG's final quality on 20 of 24 images, with a
median speed-up of \(243\times\) over those images. It reaches B\'ezier
Splatting's final quality on all 24 images, at a median \(196\times\).

The quality gap grows with shape count on \emph{F\"arlev}. At
\(512\times288\) with 4,096 shapes and 800 steps, Windfoil reaches
26.77\,dB in 10.1\,s, compared with 21.31\,dB in 1,658\,s for DiffVG
and 22.61\,dB in 132\,s for B\'ezier Splatting. With a fixed 60\,s
optimisation budget at \(4096\times2304\) and 256 shapes, Windfoil reaches
20.96\,dB, compared with 19.96\,dB for DiffVG and 19.77\,dB for B\'ezier
Splatting.

B\'ezier Splatting's published schedule uses 10,000 steps, so we also give
both it and Windfoil this budget on \emph{F\"arlev}. Windfoil leads by
0.53\,dB at 512 shapes and 4.86\,dB at 4,096 shapes, with parity speed-ups
of \(947\times\) and \(2{,}145\times\), respectively.

To test larger scenes, we also fit \emph{F\"arlev} at \(4096\times2304\)
with 10,000 and 50,000 shapes. Windfoil reaches 23.23 and 25.46\,dB in
65.6 and 176\,s for 800 steps. The baselines were not tested at these counts.

Each benchmark was run once using the same initialisation seed. Timings can vary
across the same GPU on different execution hosts due to various CPU efficiencies. 

\begin{figure*}[t]
  \centering
  \setlength{\clipgut}{2pt}
  \setlength{\clipgap}{8\clipgut}
  \setlength{\cliptilew}{\dimexpr(\linewidth-5\clipgut-4\clipgap)/10\relax}
  \begin{tabular}{@{}c@{\hspace{\clipgap}}c@{\hspace{\clipgap}}c@{\hspace{\clipgap}}c@{\hspace{\clipgap}}c@{}}
    \clipgrid{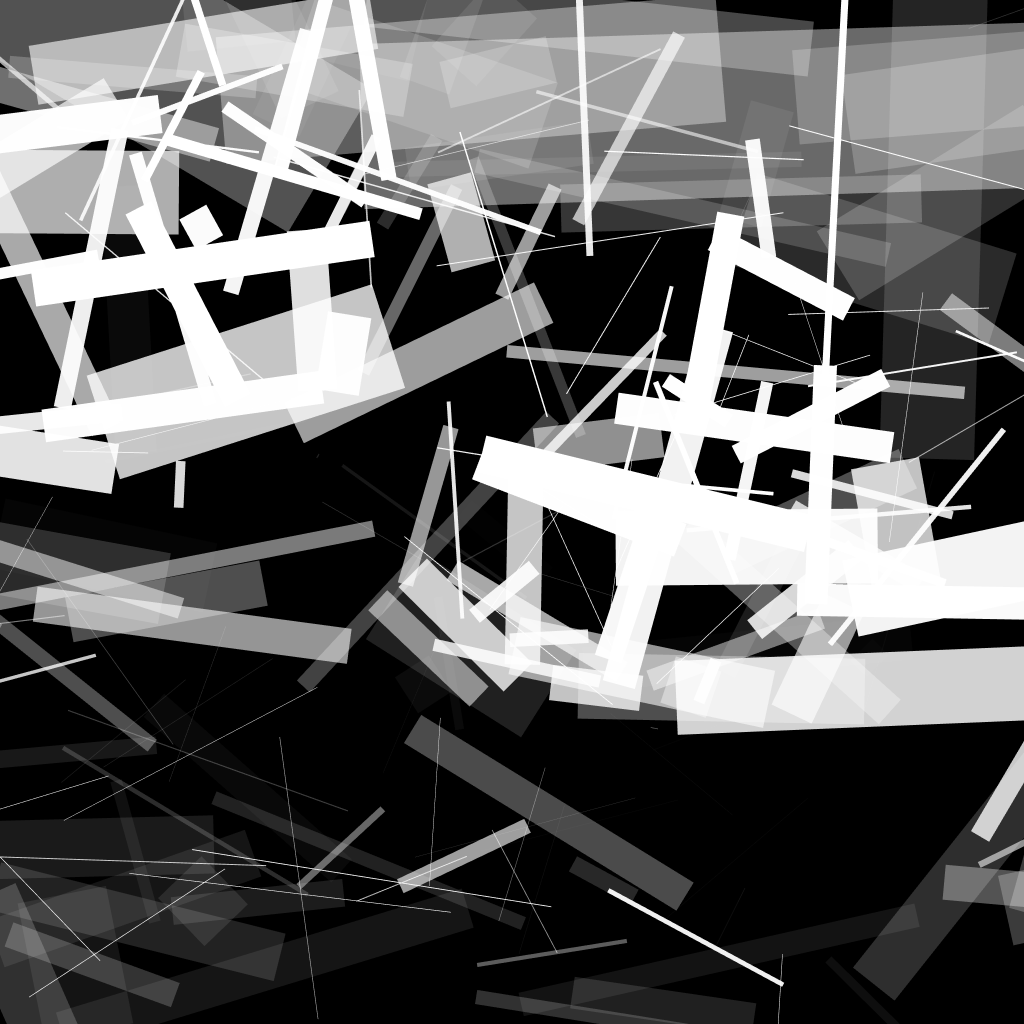}{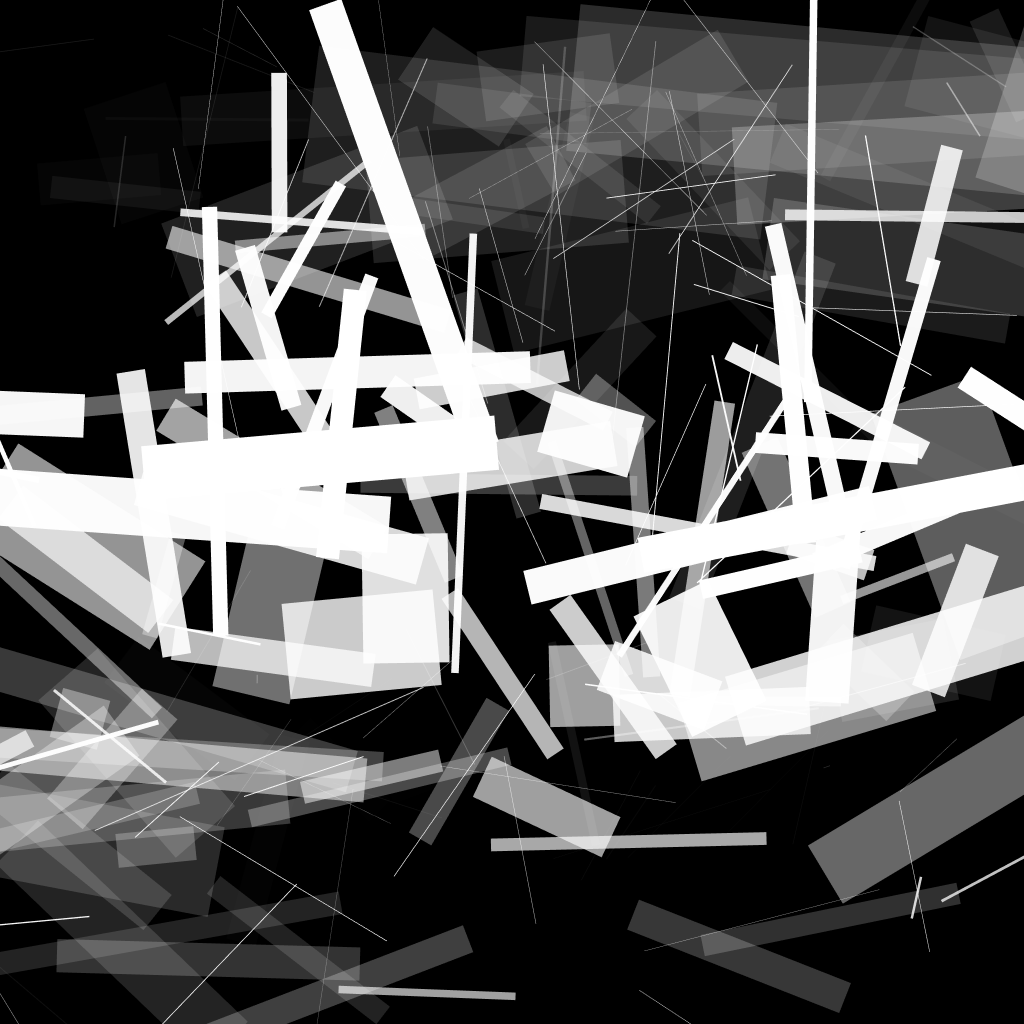}
             {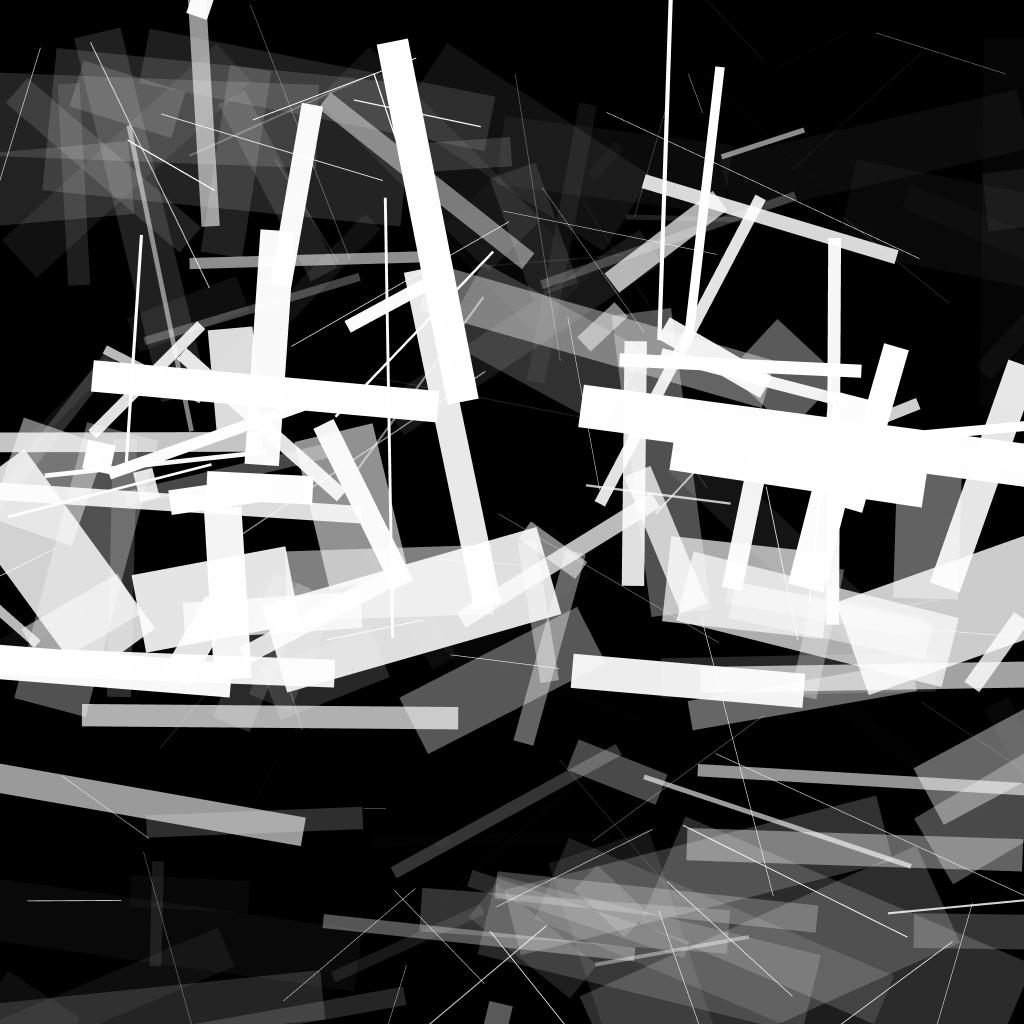}{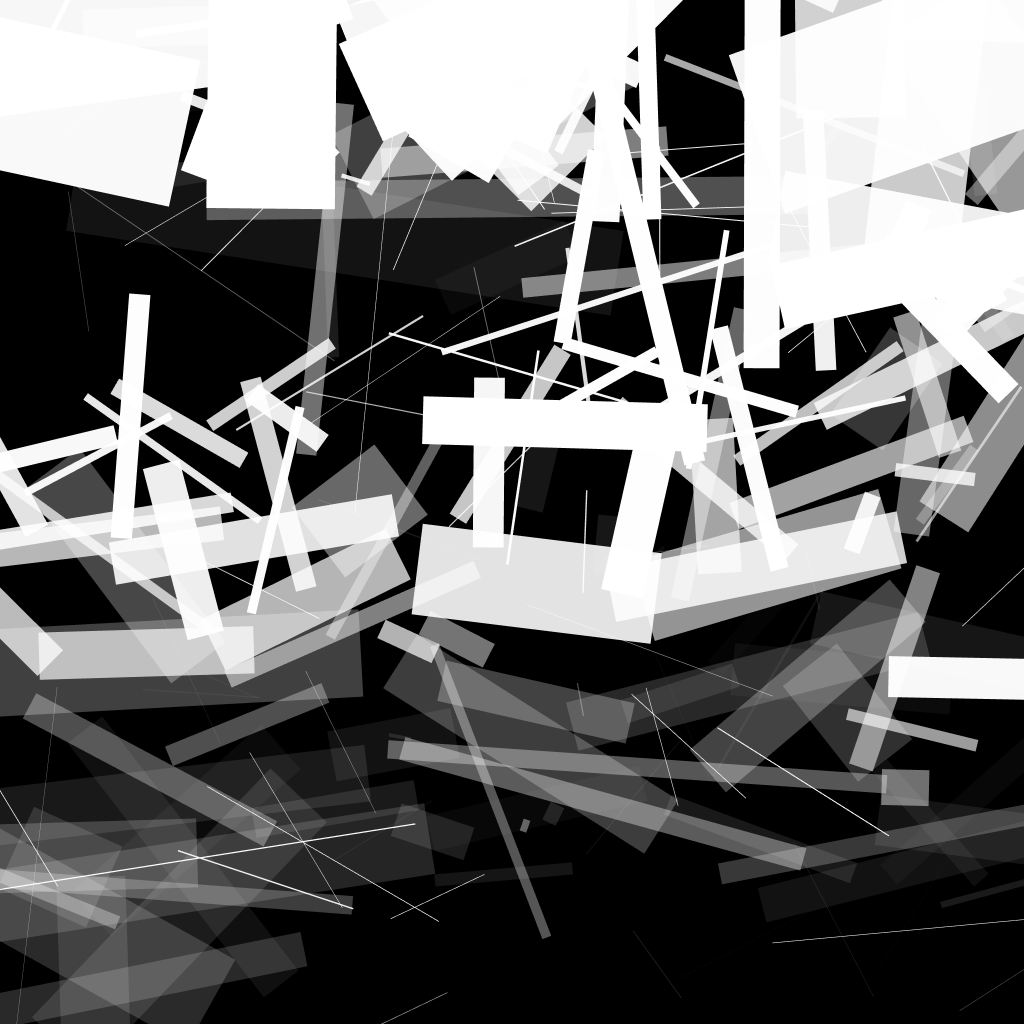} &
    \clipgrid{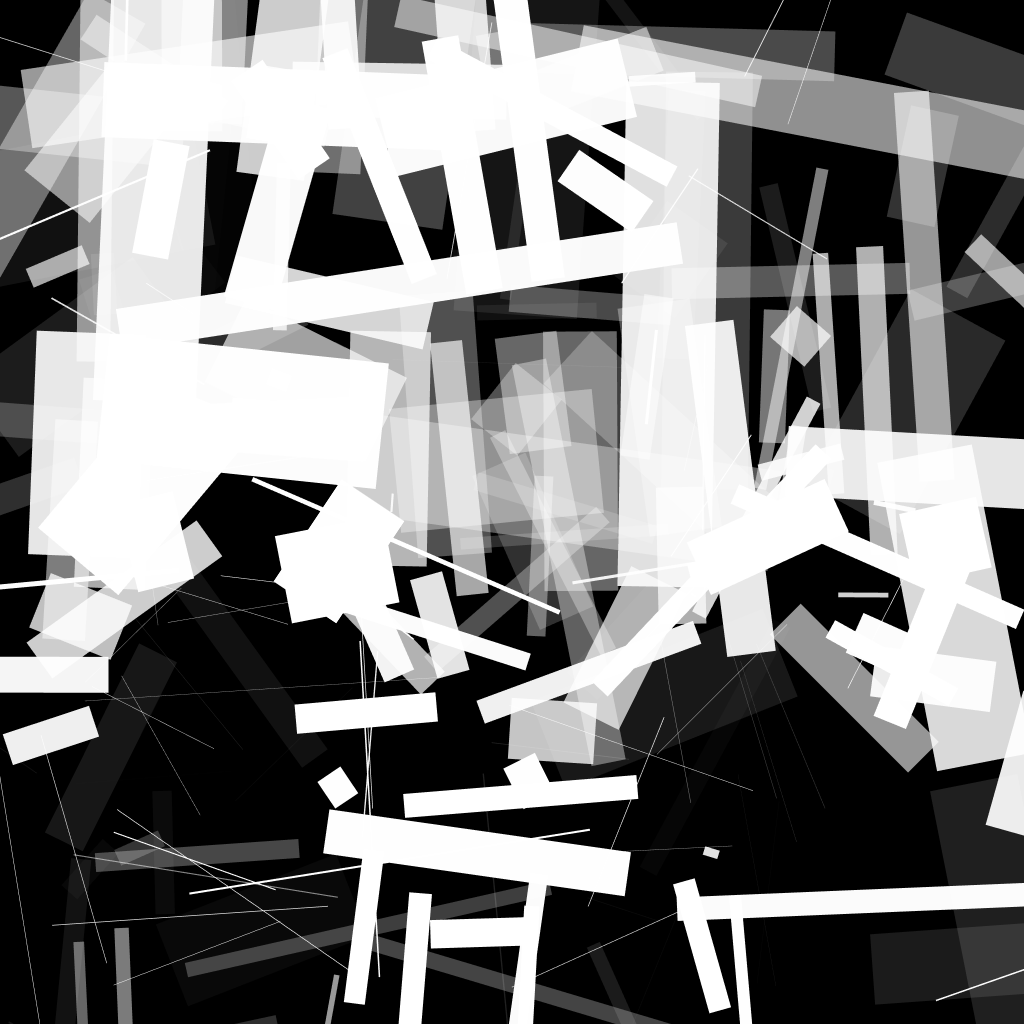}{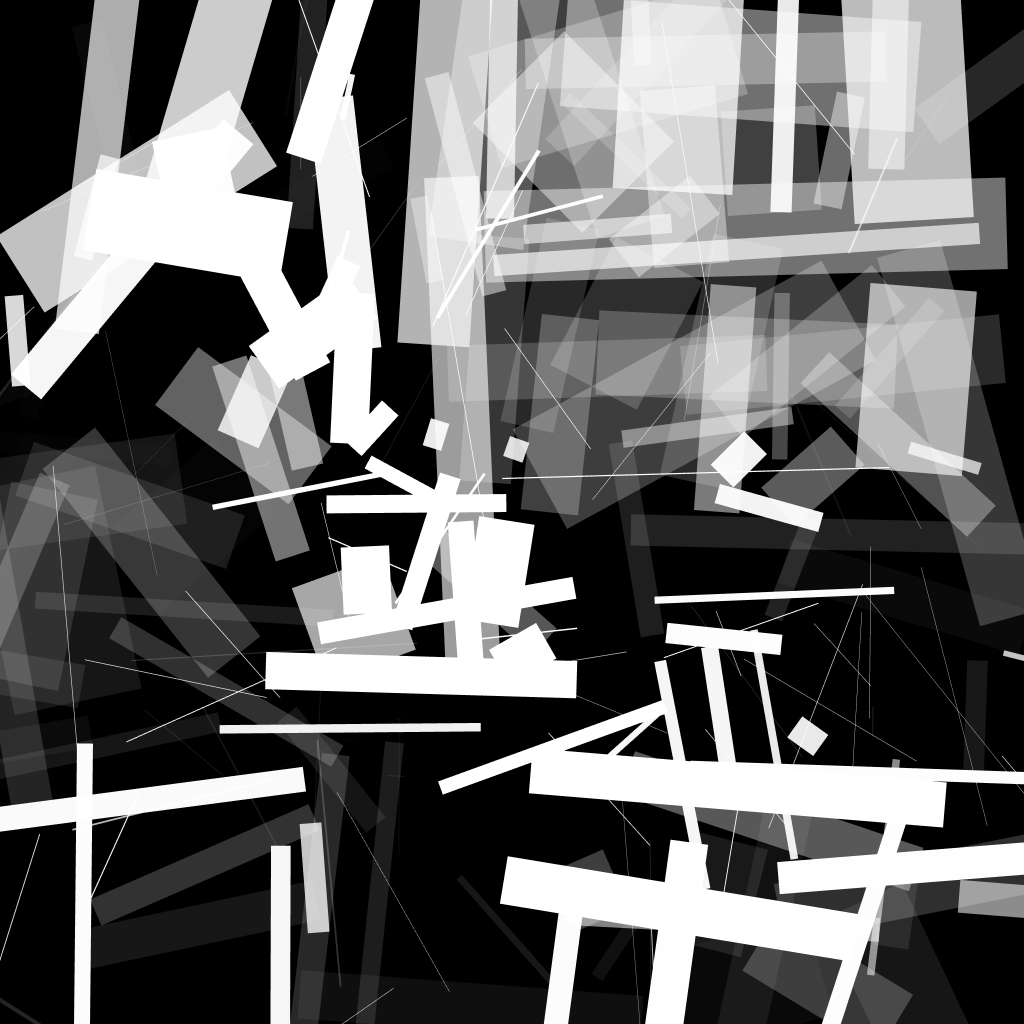}
             {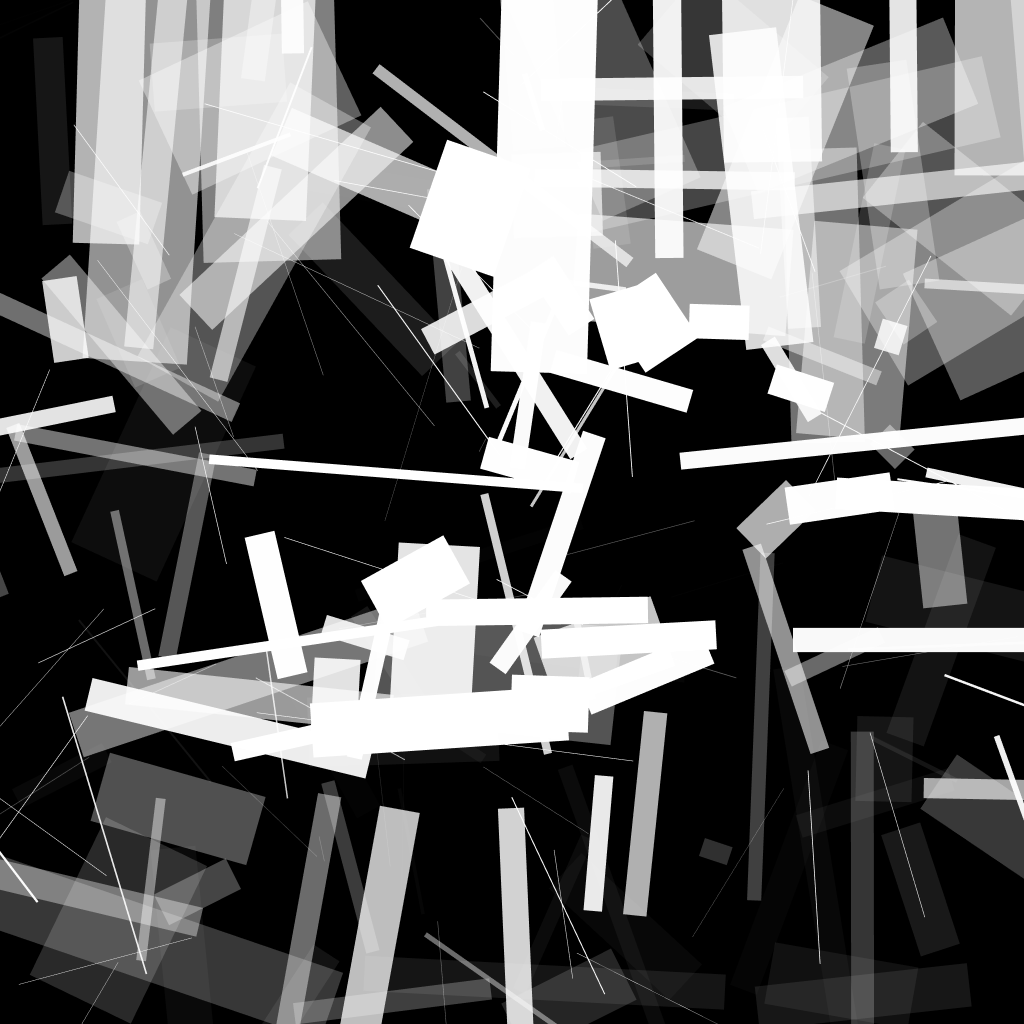}{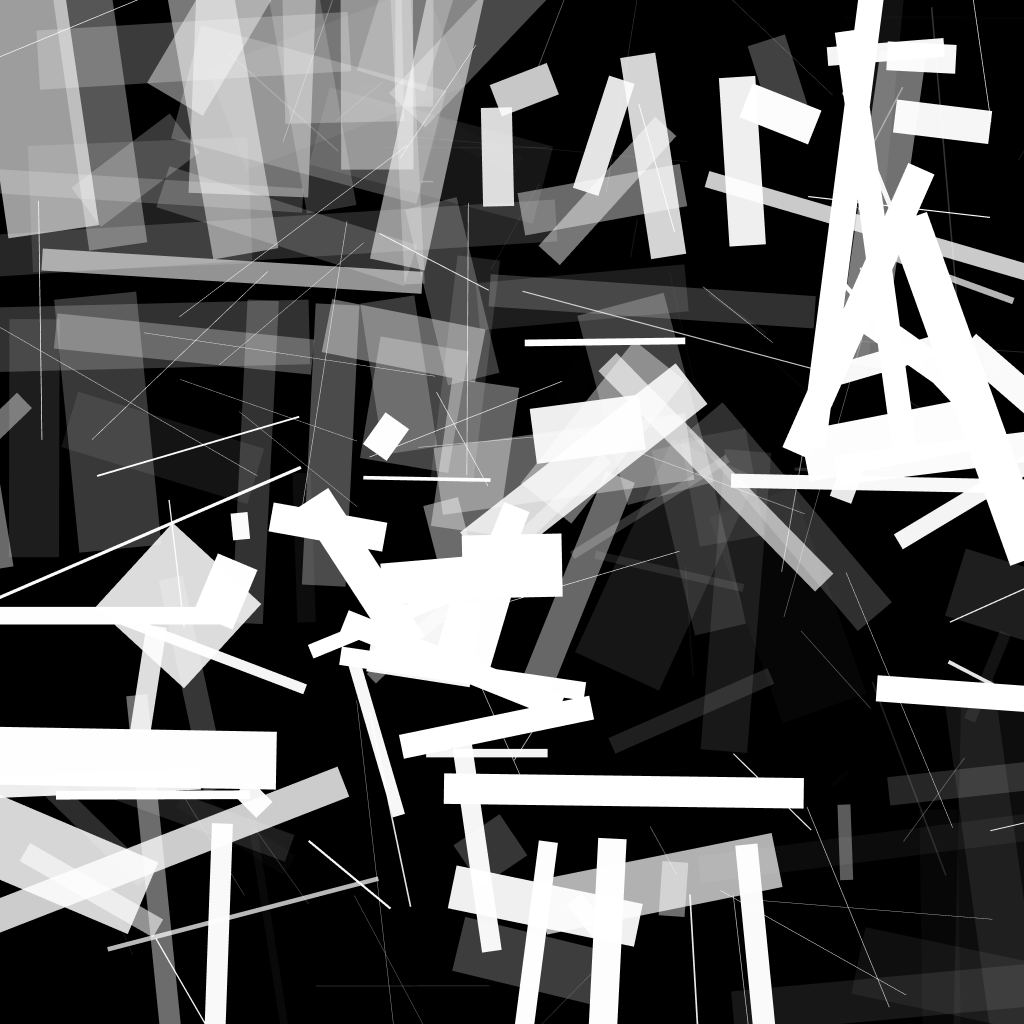} &
    \clipgrid{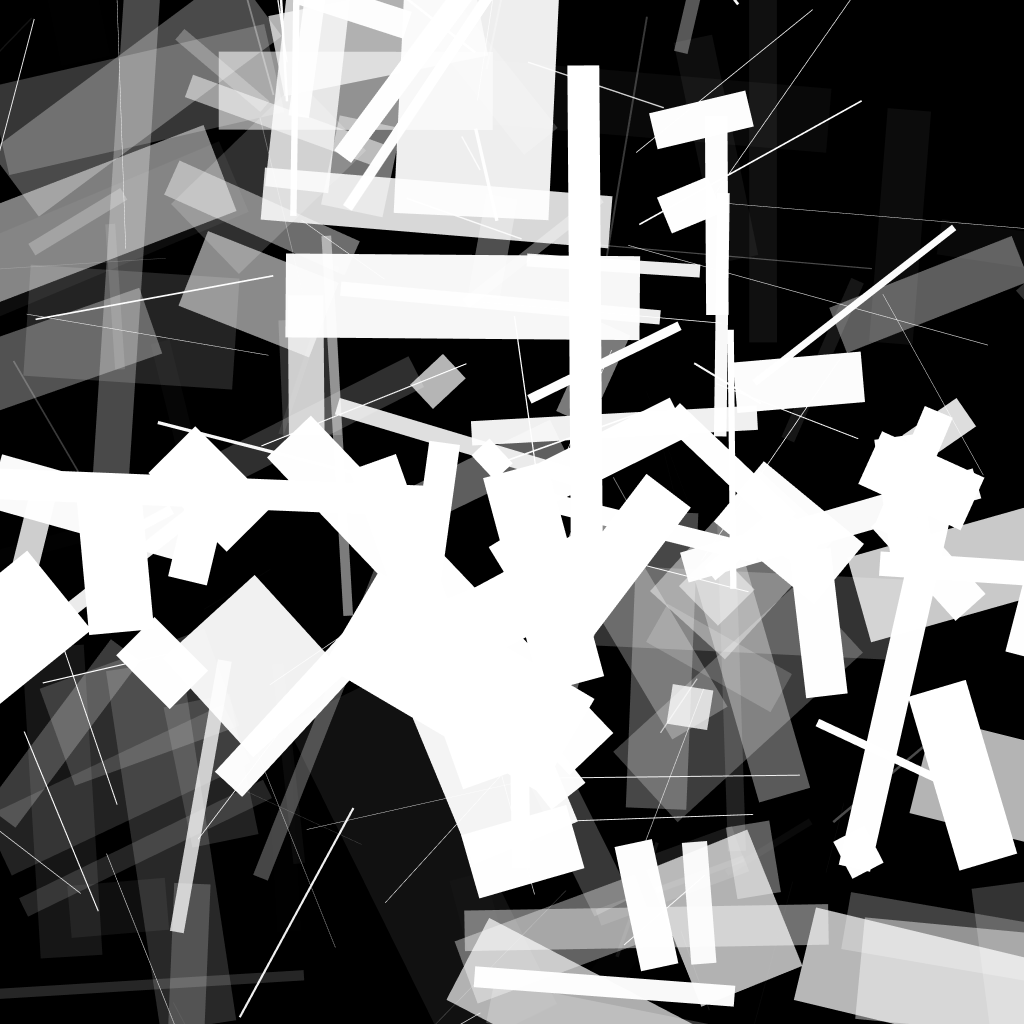}{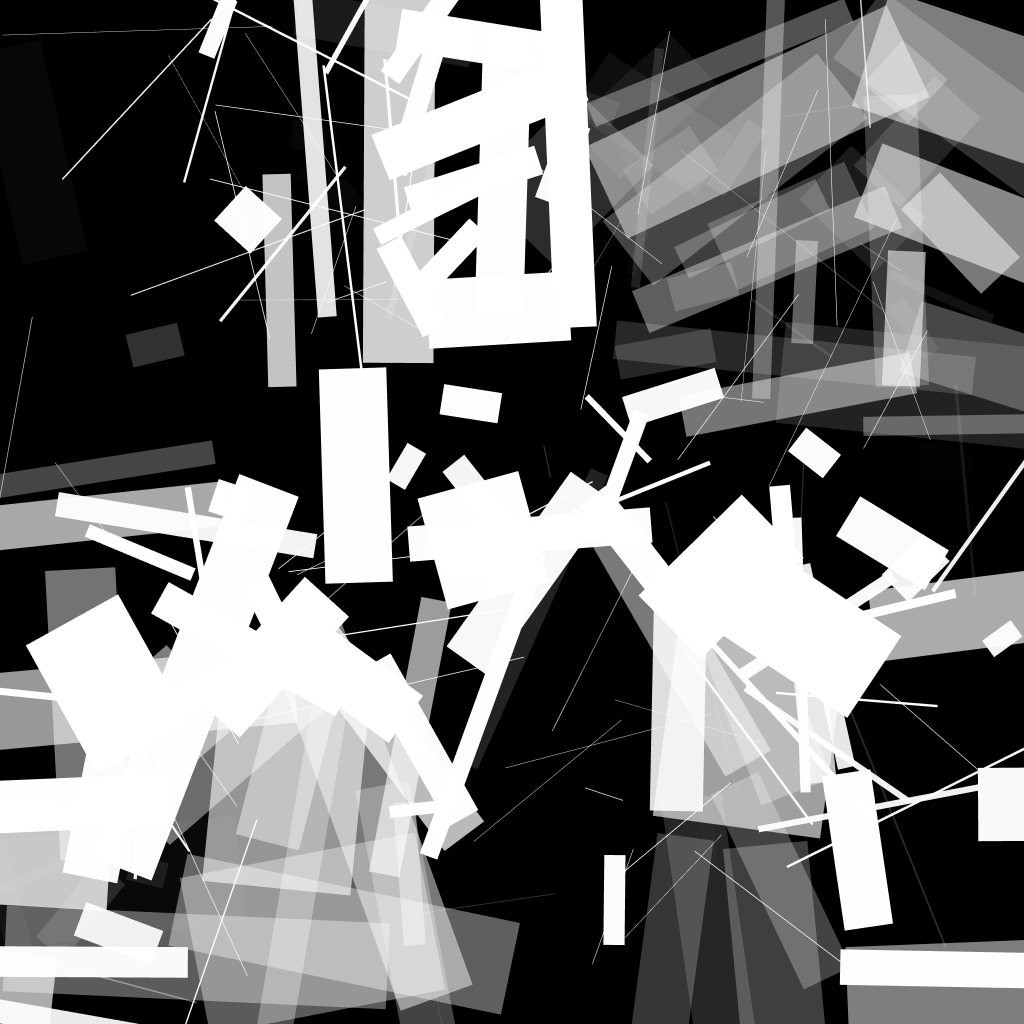}
             {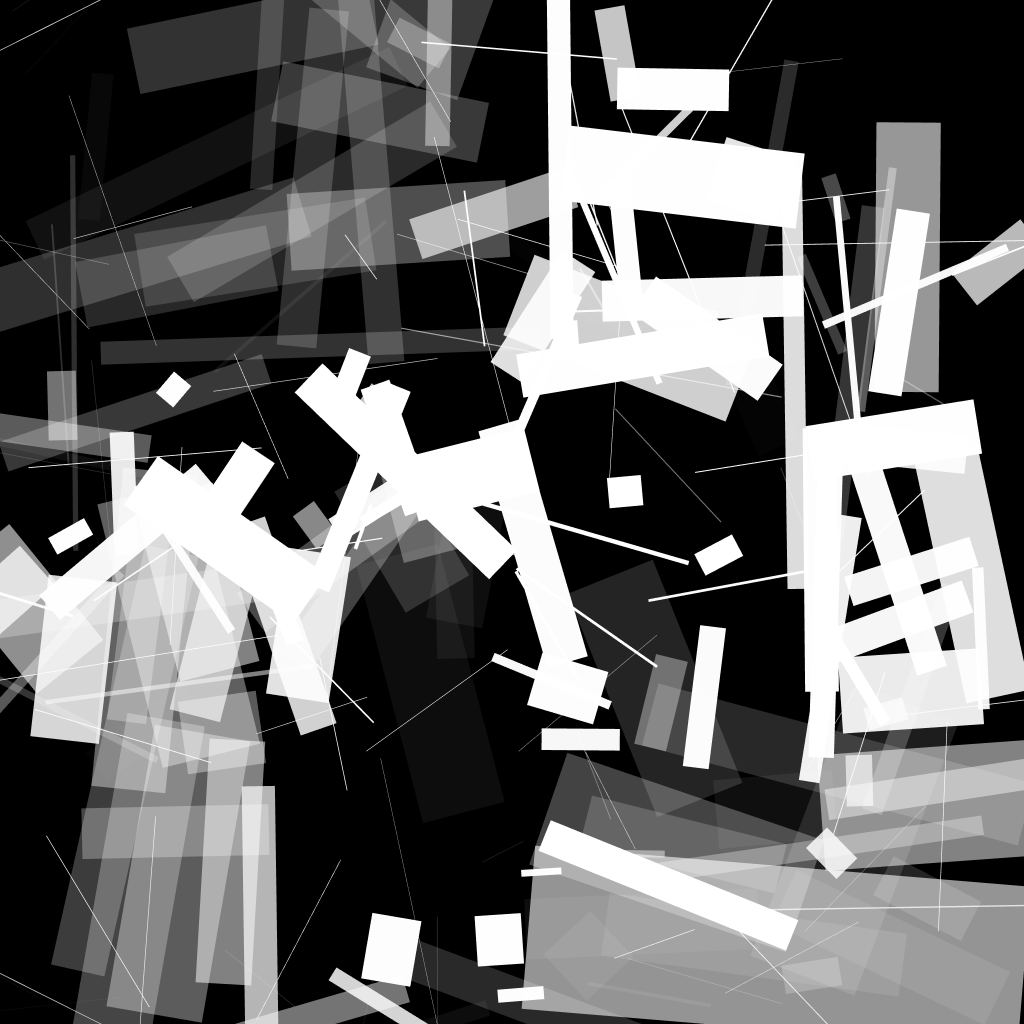}{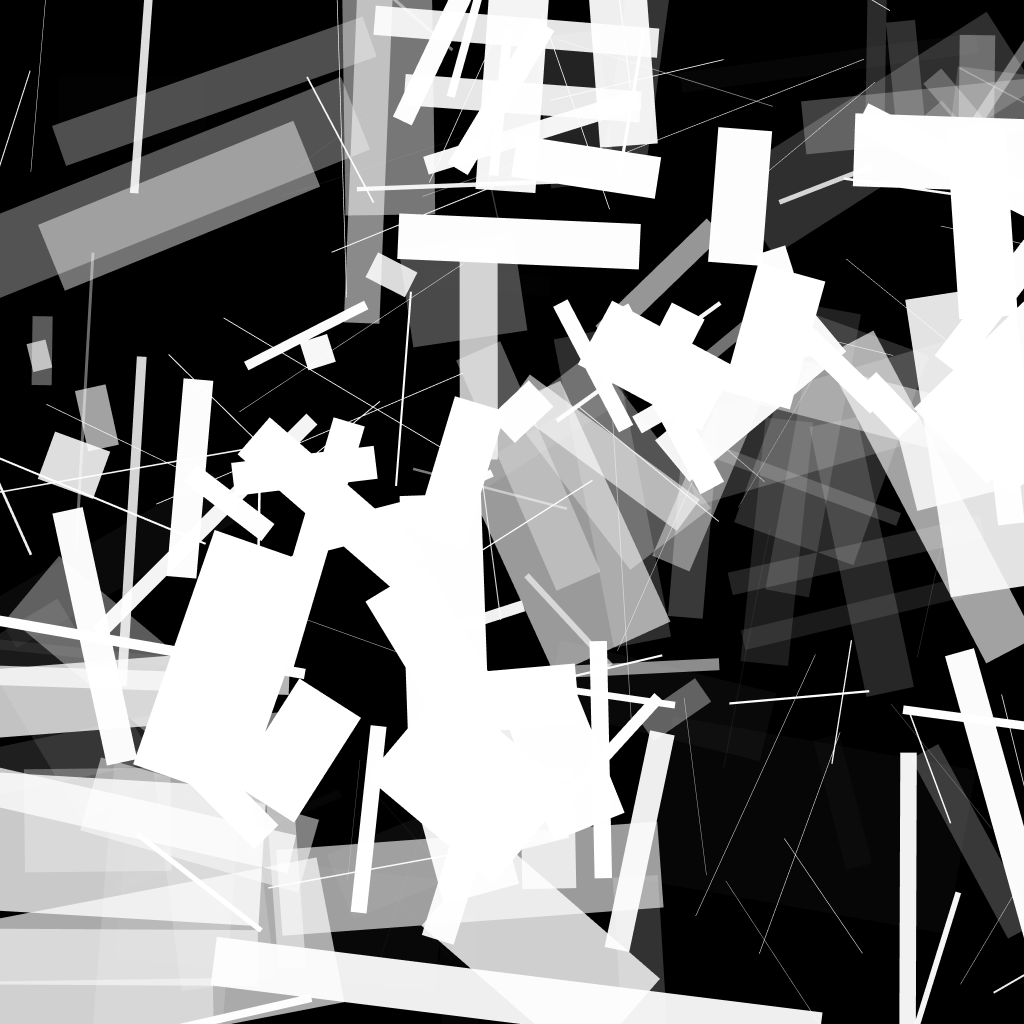} &
    \clipgrid{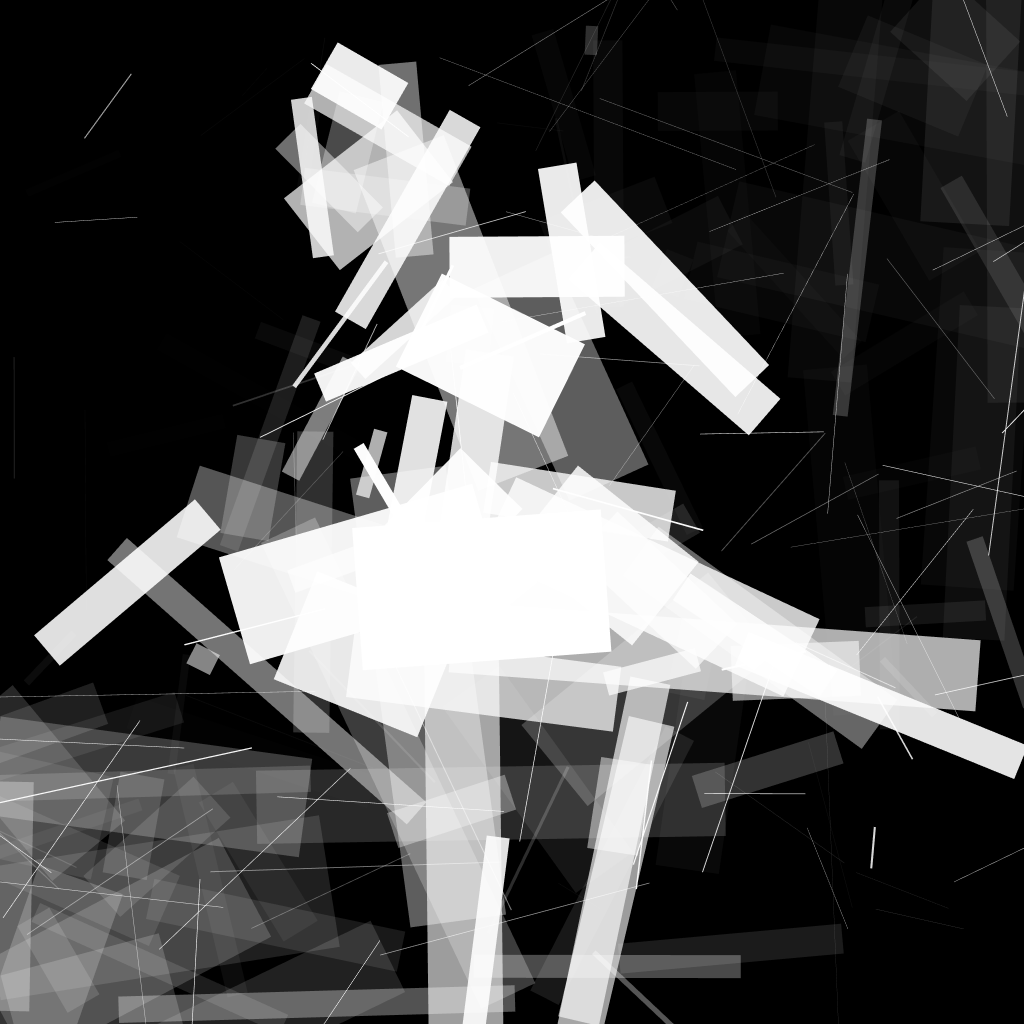}{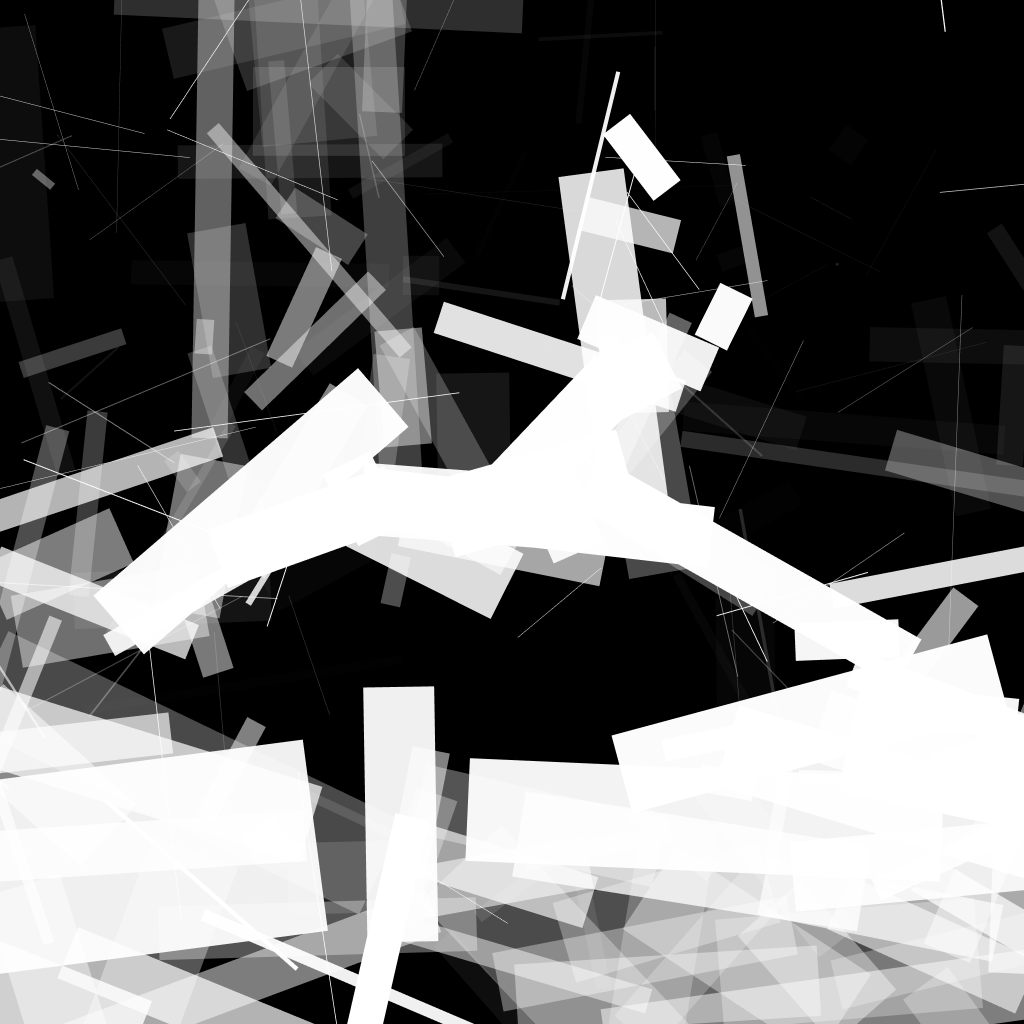}
             {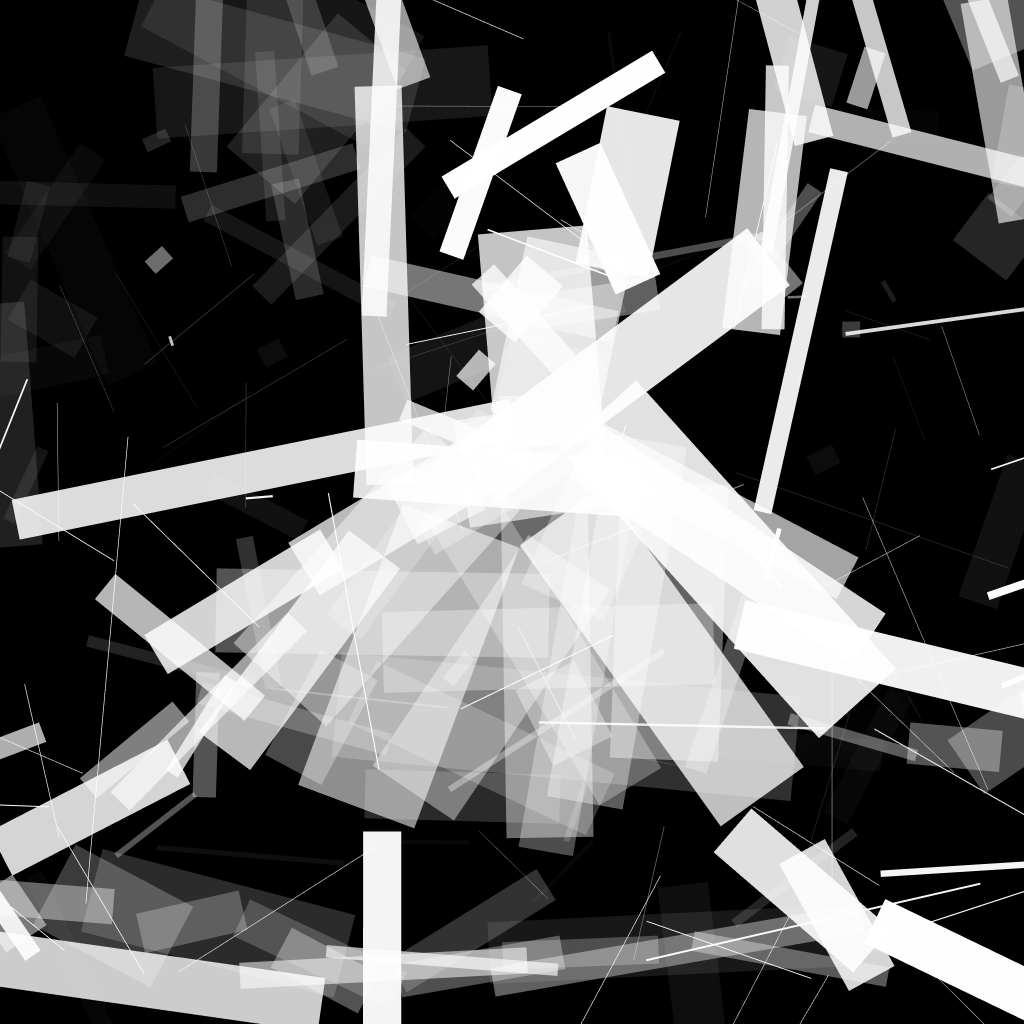}{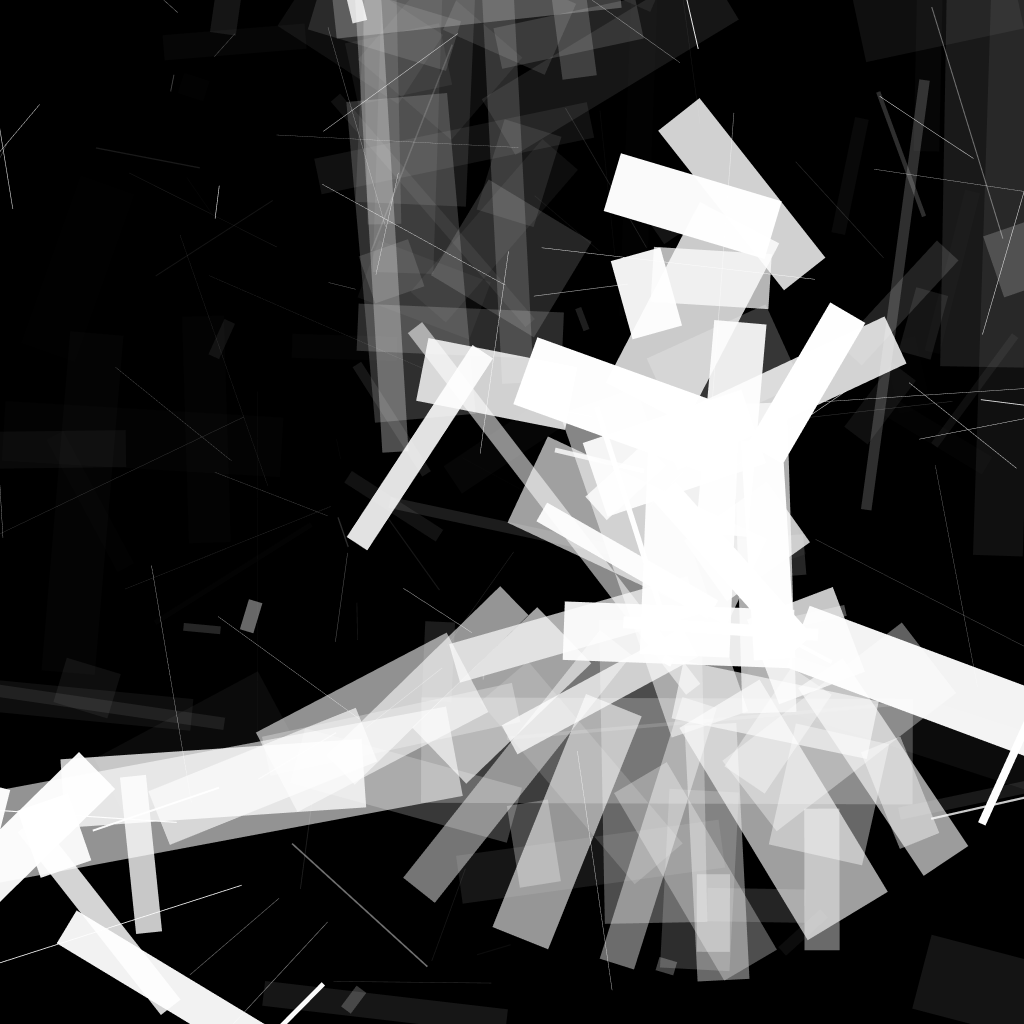} &
    \clipgrid{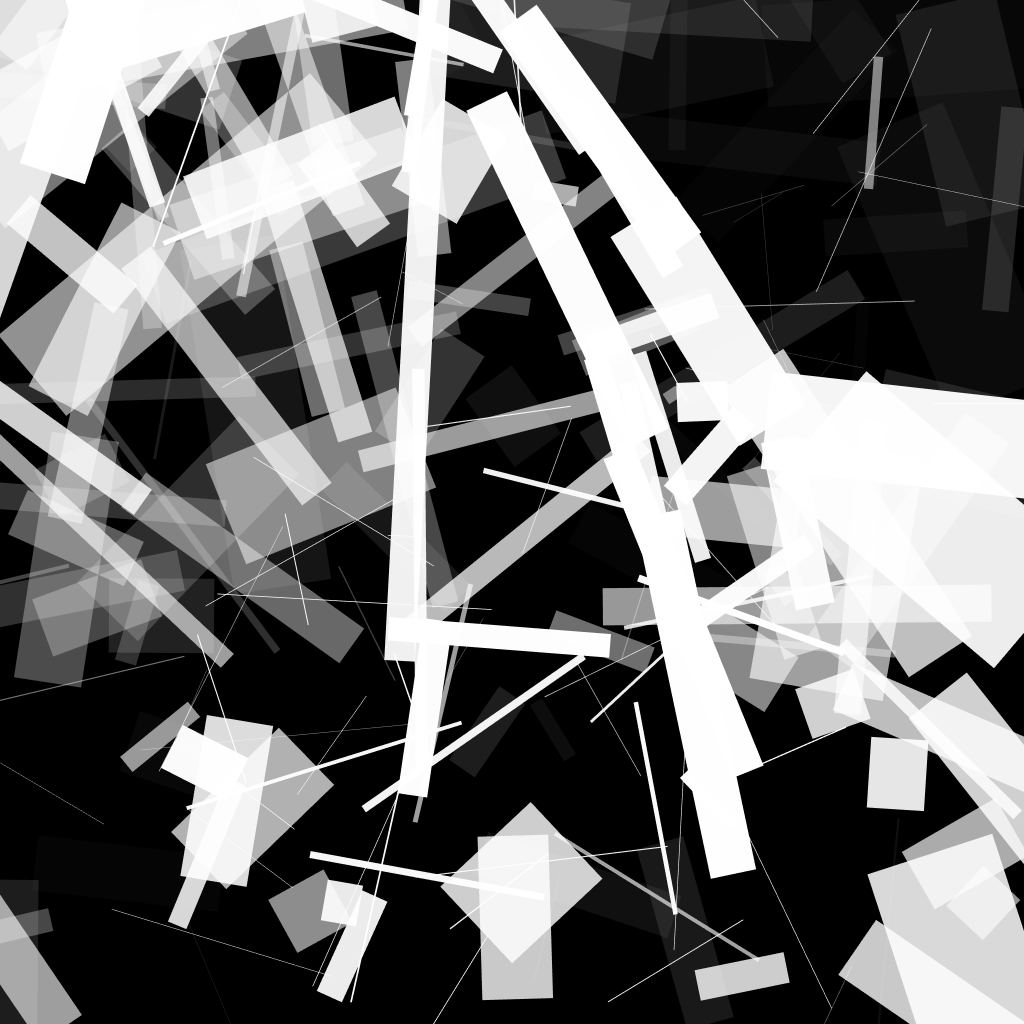}{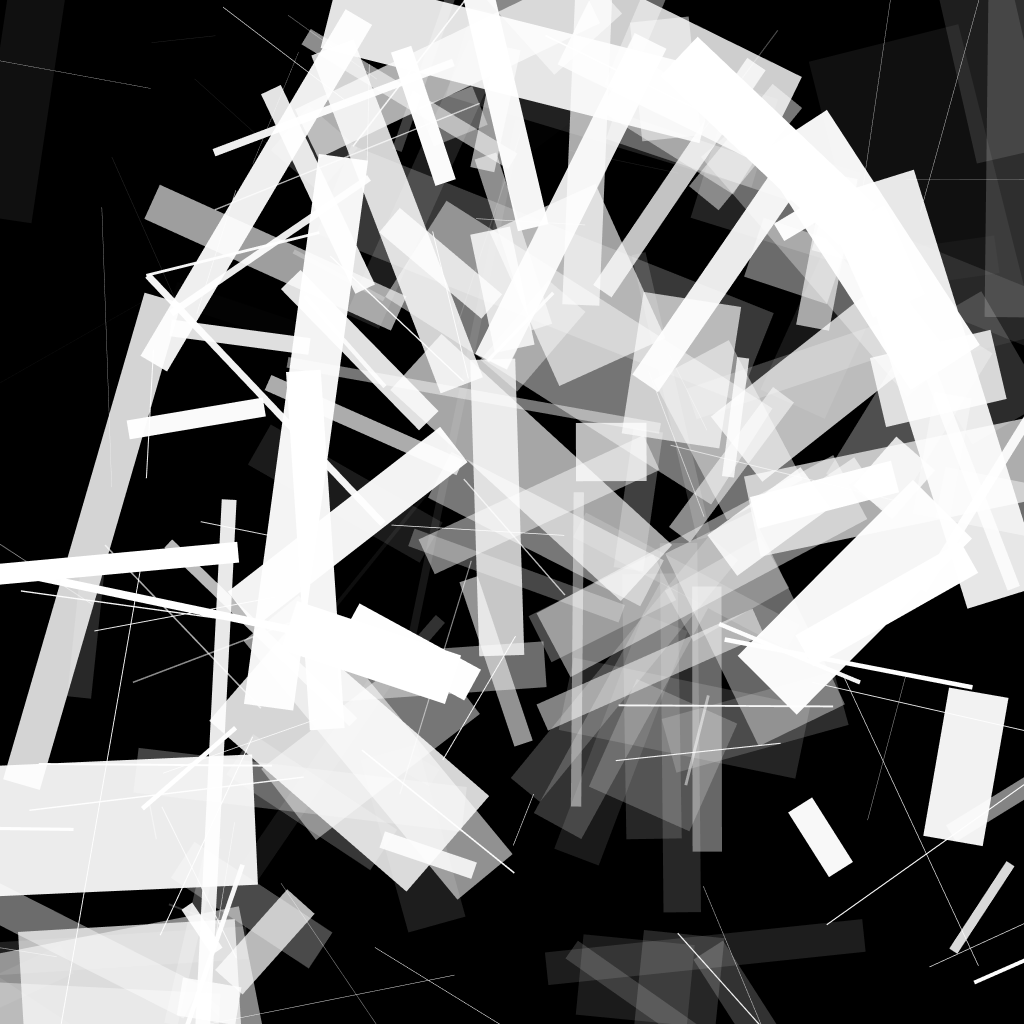}
             {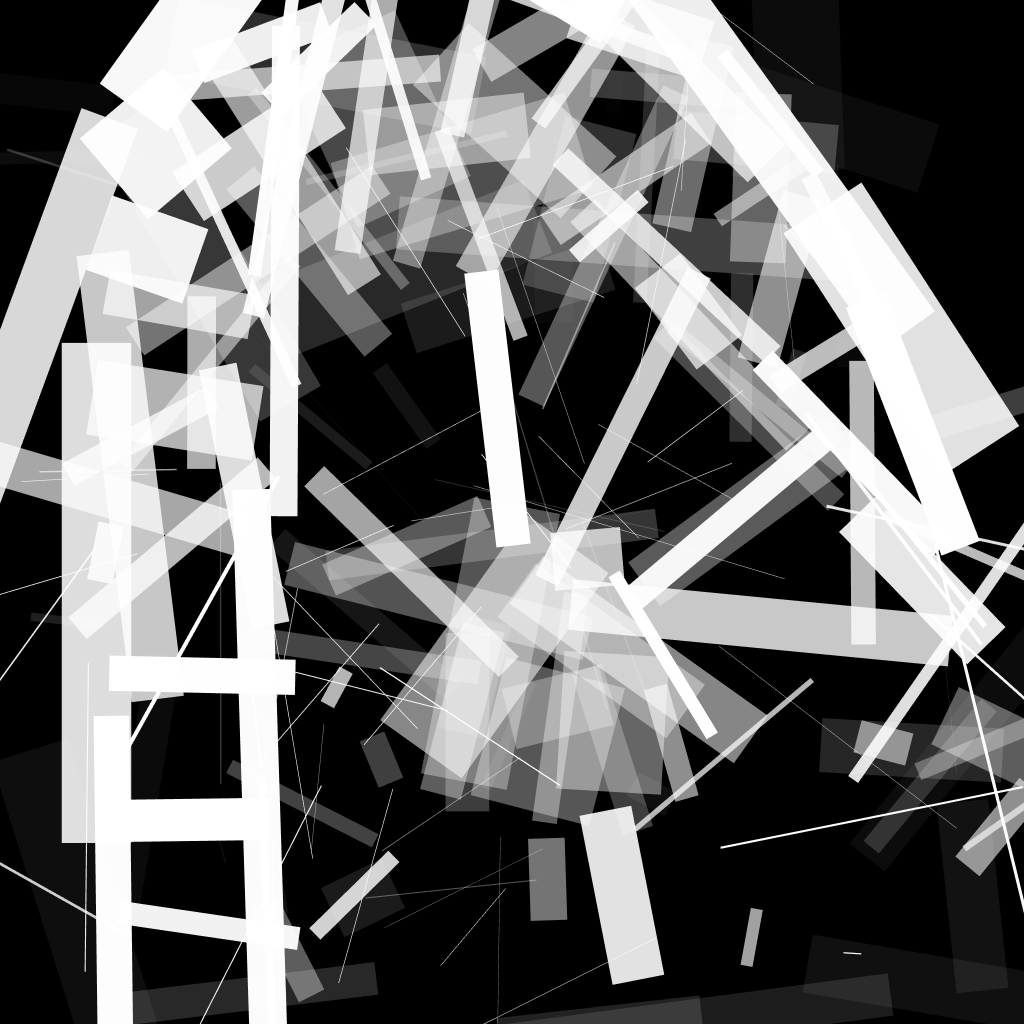}{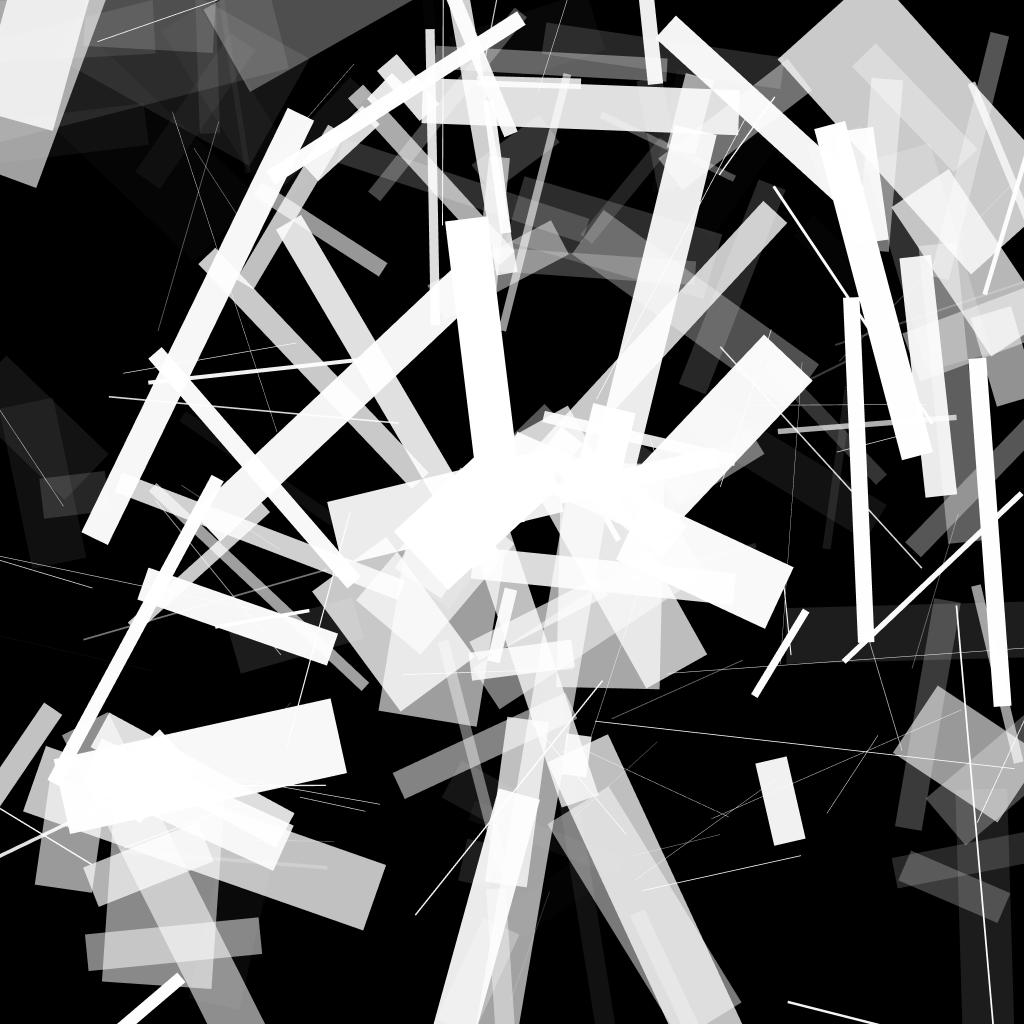} \\[2pt]
    \footnotesize \emph{fishing boats in the harbour} &
    \footnotesize \emph{a Paris caf\'e} &
    \footnotesize \emph{a Tokyo street} &
    \footnotesize \emph{a ballerina} &
    \footnotesize \emph{a Ferris wheel} \\
  \end{tabular}
  \caption{Text-to-SVG synthesis of scenes made of 256 white translucent rectangles, guided by CLIP. Each 2$\times$2 grid shows four optimisation runs using different random seeds.}
  \Description{Five two-by-two grids of black-and-white generated scenes, one
    grid per text prompt. Each sketch is built from a few hundred
    overlapping semi-transparent white shapes on a black background.}
  \label{fig:clip}
\end{figure*}

\section{Use Cases}

Below we highlight some features and applications enabled by our renderer.

\paragraph{Adjustable Filter Kernels.}
Most production 2D renderers provide little control over the filter used for
antialiasing, although experimental systems such as Olson's
Alpenglow~\cite{olson2024exact} have explored analytic filtering under several
kernels. In Windfoil, the filter size is an input to the box-filtered winding,
so we can widen it to produce an analytic box blur. This is
particularly useful during differentiable rendering, where we anneal from a
wide filter to a single pixel, fitting coarse-to-fine detail. The
boundary-integral formulation also allows us to replace the box filter with
other kernels while remaining differentiable. We demonstrate this with tent
and cubic B-spline filters using quadrature approximations
(Figure~\ref{fig:kernels}). These filters have
a notable performance cost and are best suited to reducing moir\'e and
shimmering in dense line work. We show this on a sunburst scene rendered with
each kernel and compare the result against Skia, Figma, and Photoshop.

\paragraph{High-Resolution Print Export.}
Several features of Windfoil make it attractive for high-resolution print
media. Its independent per-pixel evaluation allows us to render very large
images as smaller tiles, which are encoded and compressed in separate worker
threads without ever storing the full PNG in active GPU or CPU memory; we have
tested images up to \(30{,}000\times30{,}000\) pixels. Because coverage is not
accumulated between neighbouring pixels, these tiles can be stitched without
artefacts at tile boundaries. Adjustable filter width and kernel selection can
give artists and designers more control over the final rasterisation and help
reduce aliasing and moir\'e in thin line work. Hairline-scale geometry
is handled by the same box-filtered winding calculation as any other shape, allowing
thin features to diminish continuously and predictably without a separate hairline
renderer. Finally, compositing can be performed in floating point and can
support high-bit-depth image exports, wide colour gamuts, and high dynamic
range.

\paragraph{Image Fitting.}
We evaluate image fitting under mean-squared error as our main differentiable benchmark, and
Section~\ref{sec:results-differentiation} compares Windfoil with DiffVG and
B\'ezier Splatting (Table~\ref{tab:fitting}, Figure~\ref{fig:convergence}).
Windfoil runs across several runtimes, including a web browser, enabling real-time, interactive,
and offline applications, such as tracing vector outlines to a target image (Figure~\ref{fig:fitting}).

\paragraph{Text-to-Sketch.}
There is growing interest in using differentiable renderers to optimise vector
scenes toward a semantic objective rather than fitting a raster
target~\cite{frans2022clipdraw,vinker2022clipasso,vinker2023clipascene,
xing2023diffsketcher,xing2024svgdreamer}. We experiment with this using CLIP,
a pretrained vision-language model, to guide curve control points and
appearance (Figure~\ref{fig:clip}). A small Python loss server supplies the cosine distance between the
rendered image's embedding and a text embedding. Across a range of prompts and
scene initialisations, our approach reaches a lower CLIP loss than DiffVG at
64, 128, and 256 shapes, and we also test larger scenes with 50,000 or 100,000
shapes.

\section{AI Disclosure}

Windfoil was developed through a close collaboration with Claude Code (Fable)
and related large language model (LLM) tooling, and we believe it is important
to highlight this role explicitly. The formulation of a closed-form and GPU-friendly box-filtered
winding was first proposed by Fable during an approximately one-hour, author-directed
search for a new 2D rasterisation algorithm for real-time rendering. This investigation took
place within a vector graphics toolkit already under development by the author, allowing the
LLM to autonomously and iteratively test its ideas within an existing suite of
visual checks, performance benchmarks, and numerical validators.

The resulting formulation was subsequently implemented, tested, and validated
by the author. Although it draws on a number of established ideas, our literature review
did not identify prior work describing the same specific contribution: a closed-form
box-filtered winding integral evaluated independently at each pixel without multisampling,
with curve contributions gathered and summed within a single shader invocation,
and accelerated using a single-axis band structure.

After releasing the initial technical details and code online, the author
recognised that the closed-form evaluation could also be extended to a
differentiable vector renderer. Claude Code and agentic coding tools
remained closely involved throughout subsequent development, with proposed
methods repeatedly tested, benchmarked, refined, and rejected under the author's
direction.

\section{Discussion and Limitations}

The box-filtered winding form is fast enough to evaluate per fragment and
differentiates directly in closed form, so display and fitting can share one
algorithm rather than two separate approximations. The optimiser and display
therefore share the same coverage model: the loss observes the
same image that is rendered for real-time or print use.

Our approach has several limitations. Averaging the winding before applying a
fill rule is not generally equivalent to filtering a filled image. The two approaches
agree along ordinary boundaries, but can differ where a pixel footprint spans
certain edge cases such as self-crossing curves. The pixel footprint is an axis-aligned box,
which exactly represents a device-pixel preimage under translation and axis-aligned scaling. 
However, this is not exact under rotation, shear, or perspective, and this is an
aspect that could be improved for a more generalised differentiable vector renderer.
Filtering each shape before compositing also differs from filtering the fully composited scene where
edges meet. The box-filtered winding evaluation is closed form,
but the wider tent and cubic kernels use fixed-tap quadrature over it rather
than closed-form convolutions.

Our implementation currently supports only closed quadratic contours. Stroked
paths and cubic B\'ezier curves must first be converted to closed quadratic
outlines, and features such as clipping masks and paint effects, including
gradients and textures, are not currently supported by the differentiable
renderer. Gradients are accumulated using fixed-point integer atomics, which allows
for the sums to be independent of accumulation order, but introduces quantisation and a risk
of overflow.

\section{Conclusion}

Windfoil brings together rasterisation and differentiable vector graphics through
a shared closed-form evaluation of the box-filtered winding number. In our evaluation,
it matches a box-filter reference more closely than Slug and Skia,
while offering scale-invariant rendering, adjustable filter kernels, and performance
competitive with Slug. For image fitting, it achieves comparable or better
reconstruction quality at a small fraction of the per-step
cost of DiffVG and B\'ezier Splatting.

Its WebGPU implementation enables efficient rasterisation and optimisation across
browser and server runtimes. We believe this shared formulation provides a useful
foundation for design tools, rendering systems, and machine learning applications
of vector graphics.

\balance
\bibliographystyle{ACM-Reference-Format}
\bibliography{references}

\end{document}